\documentclass{llncs}
\usepackage{enumerate}
\usepackage{etoolbox}
\usepackage{stmaryrd}
\usepackage{amssymb,amsmath}
\usepackage{mathtools}
\usepackage{microtype}
\usepackage{pifont}
\usepackage{tikz}
\usepackage{wrapfig}
\usetikzlibrary{calc,arrows,fit,backgrounds}
\usetikzlibrary{shapes.geometric}
\usetikzlibrary{decorations.pathreplacing}

\makeatletter

\newcommand\@optsub[2]{
  \ifstrempty{#2}{%
    #1%
  }{%
    #1_{#2}%
  }%
}
\newcommand\@optsup[2]{
  {#1}%
  \ifstrempty{#2}{}{^{#2}}%
}
\newcommand\@optapp[2]{
  {#1}%
  \ifstrempty{#2}{}{(#2)}%
}

\renewcommand\phi{\varphi}
\renewcommand\emptyset{\varnothing}

\tikzset{
  mono/.style={>->},
  gnode/.style={circle,fill=black,inner sep=0mm,minimum size=2mm,font=\scriptsize,text=white},
  gedge/.style={->,>=latex},
  arlab/.style={inner sep=1pt,font=\scriptsize},
  glab/.style={inner sep=1pt,font=\scriptsize},
  hyperedge/.style={shape=rectangle,draw,inner sep=0,
    minimum width=1cm,minimum height=.4cm},
  point/.style={shape=circle,inner sep=0,fill=black,
    minimum height=1pt,minimum width=1pt}
}

\newcommand\graphbox[2][grbox]{
  \begin{pgfonlayer}{background}
    \node[fit=#2] (#1) {} ;
    \fill[black!20,rounded corners=2mm,postaction={draw,black}] 
          (#1.north west) -- (#1.north east) --
          (#1.south east) -- (#1.south west) -- cycle ;
  \end{pgfonlayer}
}

\newcommand\substy[2]{
  ($ (0,0)!#1!(1,0) + (0,#2) $)
}

\newcommand\tpl[1]{\langle #1 \rangle}
\newcommand\nats{\mathbb{N}_0}

\newcommand\arright[1][]{%
 \ifstrempty{#1}%
 {\rightarrow}%
 {\mathbin{%
     \mathchoice%
     {\xrightarrow{#1}}%
     {\scalebox{.8}[1]{$\textstyle\relbar$}{\raisebox{.23ex}{$\scriptstyle #1$}}{\shortrightarrow}}%
     {\scalebox{.8}[1]{$\scriptstyle\relbar$}{\raisebox{.15ex}{$\scriptscriptstyle #1$}}{\shortrightarrow}}%
     {\scalebox{.8}[1]{$\scriptscriptstyle\relbar$}{\raisebox{.15ex}{$\scriptscriptstyle #1$}}{\shortrightarrow}}%
 }}
}

\newcommand\arleft[1][]{%
 \ifstrempty{#1}%
 {\leftarrow}%
 {\mathchoice%
   {\xleftarrow{#1}}
   {\mathbin{{\textstyle\shortleftarrow}{\raisebox{.23ex}{$\scriptstyle #1$}}\scalebox{.8}[1]{$\textstyle\relbar$}}}
   {\mathbin{{\scriptstyle\shortleftarrow}{\raisebox{.15ex}{$\scriptscriptstyle #1$}}\scalebox{.8}[1]{$\scriptstyle\relbar$}}}
   {\mathbin{{\scriptscriptstyle\shortleftarrow}{\raisebox{.15ex}{$\scriptscriptstyle #1$}}\scalebox{.8}[1]{$\scriptscriptstyle\relbar$}}}
 }
}

\newcommand\sysR{\mathcal{R}}

\newcommand\sLab[1][]{\@optsub{\mathit{lab}}{#1}}
\newcommand\sSrc[1][]{\@optsub{\mathit{src}}{#1}}
\newcommand\sTgt[1][]{\@optsub{\mathit{tgt}}{#1}}

\newcommand\fLab[2][]{\sLab[#1](#2)}
\newcommand\fSrc[2][]{\sSrc[#1](#2)}
\newcommand\fTgt[2][]{\sTgt[#1](#2)}

\newcommand\sFlower[1][]{\@optsub{\mbox{\ding{82}}}{#1}}
\newcommand\sFlowerM[1][]{\@optsub{\mathit{fl}}{#1}}

\newcommand\fFlowerM[2][]{\sFlowerM[#1](#2)}

\makeatother

\newenvironment{theorem_for}[2]{\noindent{\bf Theorem~\ref{#1}#2}\it}{}

\newenvironment{lemma_for}[2]{\noindent{\bf Lemma~\ref{#1}#2}\it}{}

\newcommand{\short}[1]{}
\newcommand{\full}[1]{#1}

\title{Proving Termination of Graph Transformation Systems using Weighted 
  Type Graphs over Semirings\thanks{Research partially supported by
    DFG project \textsc{GaReV}.}}
\author{H.J. Sander Bruggink\inst{1} \and Barbara K\"onig\inst{2} \and
  Dennis Nolte\inst{2} \and Hans Zantema\inst{3}}
\institute{%
  GEBIT Solutions \\
  \email{sander.bruggink@gebit.de} \\
  \and Universit{\"a}t Duisburg-Essen\\ 
  \email{\{barbara\_koenig,dennis.nolte\}@uni-due.de}
  \and Technische Universiteit Eindhoven and Radboud Universiteit Nijmegen\\
  \email{h.zantema@tue.nl}}

\begin{document}
\maketitle

\begin{abstract}
  We introduce techniques for proving uniform termination of graph
  transformation systems, based on matrix interpretations for string
  rewriting. We generalize this technique by adapting it to graph
  rewriting instead of string rewriting and by generalizing to ordered
  semirings. In this way we obtain a framework which is inspired by
  the tropical and arctic type graphs of \cite{bkz14} and introduces a
  new variant of arithmetic type graphs. These type graphs can be used
  to assign weights to graphs and to show that these weights decrease
  in every rewriting step in order to prove termination. We present an
  example involving counters and discuss the implementation in the
  tool Grez.
\end{abstract}

\section{Introduction}
\label{sec:introduction}

For every computational formalism, the question of termination is one
of the most fundamental problems, consider for instance the halting
problem for Turing machines. For graph transformation systems there
has been some work on termination, but this problem has received less
attention than, e.g., confluence or reachability analysis. There are
several applications where termination analysis is essential: one
scenario is termination of graph programs, especially for programs
operating on complex data structures. Furthermore, model
transformations, for instance of UML models, usually require
functional behaviour, i.e., every source model should be translated
into a unique target model.  This requires termination and confluence
of the model transformation rules.

There is a huge body of termination results in string and term
rewriting \cite{bkv:terese} from which one can draw inspiration. Still,
adapting these techniques to graph transformation is often
non-trivial. A helpful first step is often to modify these techniques
to work with cycle rewriting
\cite{zbk:termination-cycle-rewriting,sz15}, which imagines the two
ends of a string to be glued together, so that rewriting is indeed
performed on a cycle.

In this paper we focus exclusively on uniform termination, i.e., there
is only a set of graph transformation rules, but no fixed initial graph,
and the question is whether the rules terminate on \emph{all} graphs.
All variants of the termination problem, termination on all graphs as
well as termination on a fixed set of initial graphs, are undecidable
\cite{p:gts-termination-undecidable}.

In \cite{bkz14} we have shown how to adapt methods from string
rewriting
\cite{z:termination-term-semantic-labelling,kw:arctic-termination} and
to develop a technique based on weighted type graphs, which was
implemented in the tool Grez. Despite its simplicity the method is
quite powerful and finds termination arguments also in cases which are
difficult for human intuition. However, there are some examples (see
for instance the example discussed in Section~\ref{sec:example}) where
this technique fails. The corresponding techniques in string rewriting
can be seen as matrix interpretations of strings in certain semirings,
more specifically in the tropical and arctic semiring. Those semirings
can be replaced by the arithmetic semiring (the natural numbers with
addition and multiplication) in order to obtain a powerful termination
analysis method for string rewriting
\cite{hw:matrix-interpretations-string,ewz:matrix-interpretations}.

Here we generalize this method to graphs. Due to their non-linear
nature, we have to abandon matrices and instead state a different
termination criterion that is based on weights of morphisms of the
left-hand and right-hand sides of rules into a type graph. Type graphs
\cite{cmr:graph-processes} are a standard tool for typing graph
transformation systems, but we are not aware of any case where they have been 
used for termination analysis before \cite{bkz14}.

By introducing weighted type graphs we generalize matrix
interpretations for string rewriting in two ways: first, we transform
graphs instead of strings and second, we consider general semirings.
Our techniques work for so-called strictly and strongly ordered
semirings, which have to be treated in a slightly different way. 
After introducing the theory we will discuss an extended example,
followed by a presentation of the implementation in the termination
tool Grez.\footnote{\texttt{http://www.ti.inf.uni-due.de/research/tools/grez/}}
\full{All proofs can be found in Appendix~\ref{sec:proofs}.}
\short{All proofs can be found in \cite{bknz:termination-gts-semiring-arxiv}.}

\full{\paragraph*{Addendum:} The original version of this paper
  contained an error, which meant that the proofs of
  Lemmas~\ref{lem:decreasing} and~\ref{lem:decreasing-nonstrict} were
  flawed, making Theorems~\ref{thm:terminating}
  and~\ref{thm:terminating-weak} incorrect as well. This was caused by
  an inadequate definition of strongly and strictly ordered semirings
  and the fact that we allowed all kinds of weights, including
  negative weights. These problems, including counterexamples, are
  explained in detail in
  \cite{eo:generalized-weighted-termination-gts} and we would like to
  express many thanks to J{\"o}rg Endrullis and Roy Overbeek for
  making us aware of these issues, enabling us to correct them in the
  present version.}

\section{Preliminaries}
\label{sec:preliminaries}

\subsection{Graphs and Graph Transformation}
\label{sec:gts}

We first introduce graphs, morphisms, and graph transformation, in
particular the double pushout approach \cite{cmrehl:algebraic-approaches}.  In 
the context of this paper we use edge-labeled, directed graphs, but it
is straightforward to generalize the results to hypergraphs.

\begin{definition}[Graph]
  Let $\Lambda$ be a fixed set of edge labels. A
  \emph{$\Lambda$-labeled graph} is a tuple 
    $G = \tpl{V,E,\sSrc,\sTgt,\sLab}$,
  where 
    $V$ is a finite set of nodes, 
    $E$ is a finite set of edges, 
    $\sSrc,\sTgt\colon E\to V$ assign to each edge a source and
      a target, and 
    $\sLab\colon E\to\Lambda$ is a labeling function.
\end{definition}

As a notational convention, we will denote, for a given graph $G$,
its components by $V_G$, $E_G$, $\sSrc[G]$, $\sTgt[G]$ and
$\sLab[G]$, unless otherwise indicated.

\begin{definition}[Graph morphism]
  Let $G,G'$ be two $\Lambda$-labeled graphs. A \emph{graph morphism}
    $\varphi\colon G\to G'$
  consists of two functions 
    $\varphi_V\colon V_G\to V_{G'}$ and 
    $\varphi_E\colon E_G\to E_{G'}$,
  such that for each edge $e\in E_G$ it holds that 
    $\fSrc[G']{\varphi_E(e)} = \phi_V(\fSrc[G]{e})$,
    $\fTgt[G']{\varphi_E(e)} = \phi_V(\fTgt[G]{e})$ and 
    $\fLab[G']{\varphi_E(e)} = \fLab[G]{e}$.
\end{definition}

We will often drop the subscripts $V,E$ and simply write $\varphi$
instead of $\phi_V$, $\phi_E$.  We work with standard double-pushout
(DPO) graph transformation \cite{cmrehl:algebraic-approaches}. Note
that our termination results would still hold if we restricted to
injective matches. However, we can not verify instances where
termination depends on injectivity of matches (for this see
\cite{eo:generalized-weighted-termination-gts}).

\begin{definition}[Graph transformation]
  A \emph{graph transformation rule} $\rho$ consists of two morphisms
  $L \arleft[\varphi_L] I \arright[\varphi_R] R$, consisting of the
  \emph{left-hand side} $L$, the \emph{right-hand side} $R$ and the
  \emph{interface} $I$. We require that $I$ is discrete, i.e., it
  consists only of nodes.

  A \emph{match} of a left-hand side in a graph $G$ is a
  morphism $m\colon L\to G$.
  \begingroup
  Given a rule $\rho$ and a match $m\colon L\to G$, a graph $H$ is the
  result of applying \unskip\parfillskip 0pt\par
  \endgroup

  \noindent
  \begin{tabular*}{\linewidth}[t]%
      {@{}p{.7\linewidth}@{\extracolsep{\fill}}c@{\extracolsep{\fill}}r@{}}
    the rule at the match, written 
    $G \Rightarrow_{m,\rho} H$ (or $G \Rightarrow_\rho H$ if $m$ is arbitrary
    or clear from the context), if there exists a graph $C$ and 
    morphisms such that the two squares in the diagram on the right are
    pushouts in the category of graphs and graph morphisms.
  &&  
      \begin{tikzpicture}[x=1.2cm,y=-1.2cm,baseline=(i.north)]
        \node (l) at (-1,0) { $L$ } ;
        \node (i) at ( 0,0) { $I$ } ;
        \node (r) at ( 1,0) { $R$ } ;
        \node (g) at (-1,1) { $G$ } ;
        \node (c) at ( 0,1) { $C$ } ;
        \node (h) at ( 1,1) { $H$ } ;
        \draw[->] (i) -- node[arlab,above] {$\phi_L$} (l) ;
        \draw[->] (i) -- node[arlab,above] {$\phi_R$} (r) ;
        \draw[->] (l) -- node[arlab,left] {$m$} (g) ;
        \draw[->] (i) -- (c) ;
        \draw[->] (r) -- (h) ;
        \draw[->] (c) -- (g) ;
        \draw[->] (c) -- (h) ;
        \node at (-.5,.5) { \textsc{(po)} } ;
        \node at ( .5,.5) { \textsc{(po)} } ;
      \end{tikzpicture}
  \end{tabular*}
  
  \smallskip

  \indent A \emph{graph transformation system} $\sysR$ is a finite set
  of graph transformation rules.  For a graph transformation system
  $\sysR$, $\Rightarrow_{\sysR}$ is the rewriting relation on graphs
  induced by those rules.
\end{definition}

Intuitely in a graph transformation step from $G$ to $H$, the images
of all elements of the left-hand side $L$, which are not present in
the interface $I$ are deleted, and the right-hand side $R$ is added,
by gluing it to the interface.

Although the graph transformation systems themselves are untyped, our
method for termination analysis is based on type graphs
\cite{cmr:graph-processes}. For given graphs $G,T$, where $T$ is
considered as a \emph{type graph}, we say that $G$ is \emph{typed
  over} $T$ whenever there is a morphism $t\colon G\to T$. The
morphism $t$ will also be called \emph{typing morphism}. We need a
way to compose and decompose typing morphisms.

\newcommand{\lemmediating}{Let a pushout $\mathit{PO}$ consisting of objects
  $G_0,G_1,G_2,G$ be given.
  Then there exists a bijection between pairs of commuting morphisms 
    $t_1\colon G_1\to T$,
    $t_2\colon G_2\to T$
  and morphisms $t\colon G\to T$ (see diagram below). 
  \begin{center}
    \begin{tikzpicture}[x=1.1cm,y=-1.1cm,baseline=(g0)]
      \node (g0) at (0, 0) {$G_0$} ;
      \node (g1) at (1,-1) {$G_1$} ;
      \node (g2) at (1, 1) {$G_2$} ;
      \node (g)  at (2, 0) {$G$}   ;
      \node (t)  at (3, 0) {$T$}   ;
      \draw[->] (g0) -- node[arlab,above left] {$\psi_1$} (g1) ;
      \draw[->] (g0) -- node[arlab,below left] {$\psi_2$} (g2) ;
      \draw[->] (g1) -- node[arlab,above right] {$\phi_1$} (g) ;
      \draw[->] (g2) -- node[arlab,below right] {$\phi_2$} (g) ;
      \draw[->] (g)  -- node[arlab,above] {$t$} (t) ;
      \draw[->] (g1) to[bend left] node[arlab,above] {$t_1$} (t) ;
      \draw[->] (g2) to[bend right] node[arlab,below] {$t_2$} (t) ;
      \node at (1,0) { \textsc{(po)} } ;
    \end{tikzpicture}
  \end{center}
  For each $t$ we obtain a unique pair of morphisms $t_1,t_2$ by
  composing with $\phi_1$ and $\phi_2$, respectively.  Conversely, for
  each pair $t_1,t_2$ of morphisms with $t_1\circ\psi_1 =
  t_2\circ\psi_2$ we obtain a unique $t\colon G\to T$ as mediating
  morphism. In this case we will write
  $\mathit{med}_{\mathit{PO}}(t_1,t_2) = t$ and
  $\mathit{med}^{-1}_{\mathit{PO}}(t) = \tpl{t_1,t_2}$.
}

\begin{lemma}
  \label{lem:med}
  \lemmediating
\end{lemma}

\subsection{Matrix Interpretations for String Rewriting}
\label{sec:matrix-interpretations}

Our technique is strongly influenced by matrix interpretations for
proving termination in string, cycle and term rewriting systems
\cite{hw:matrix-interpretations-string,sz15,ewz:matrix-interpretations}.
We will generalize this technique, resulting in a technique for graph
transformation systems that has a distinctly different flavour than
the original method. In order to point out the differences later and
motivate our choices, we will introduce matrix interpretations first.

We are working in the context of string rewrite systems, where a rule
is of the form $\ell\to r$, where $\ell,r$ are both strings over a
given alphabet $\Sigma$. For instance, consider the rule $aa\to aba$,
which rewrites $aaa \Rightarrow abaa \Rightarrow ababa \not\Rightarrow$.

We first start with some preliminaries: let $A,B$ be two square
matrices $A,B$ over $\nats$ of equal dimension $n$. We write $A > B$
if $A_{1,1} > B_{1,1}$ and $A_{i,j} \ge B_{i,j}$ for all indices $i,j$
with $1\le i,j \leq n$, i.e., we require that the entries in the upper
left corner are strictly ordered, whereas the remaining entries may
also be equal.  It holds that $A>B$ implies $A\cdot C > B\cdot C$ and
$C\cdot A> C\cdot B$ for a matrix\footnote{Here $0$ denotes the matrix
  with all entries zero.} $C > 0$ of appropriate dimension.

As is standard in termination analysis strings are assigned to
elements in a well-founded set and it has to be shown that each rule
application leads to a decrease within this order.

Here, every letter of the alphabet $a\in\Sigma$ is associated with a
square matrix $A = [a] > 0$ (where all matrices have the same
dimension~$n$). Similarly every word $w = a_1\dots a_n$ is mapped to a
matrix $[w] = [a_1]\cdot \dots \cdot[a_n]$, which is obtained by
taking the matrices of the single letters and multiplying them. If we
can show $[\ell] > [r]$ for every rule $\ell\to r$, then termination
is implied by the considerations above and by the fact that the order
$\le$ on $\nats$ is well-founded, i.e., there are no infinite strictly
decreasing chains.

For the example above take the following matrices (as in
\cite{hw:matrix-interpretations-string}):
\[ [a] =
\begin{pmatrix}
  1 & 1 \\
  1 & 0 
\end{pmatrix}
\qquad
[b] =
\begin{pmatrix}
  1 & 0 \\
  0 & 0
\end{pmatrix}
\qquad\mbox{with}\qquad
[aa] =
\begin{pmatrix}
  2 & 1 \\
  1 & 1 
\end{pmatrix}
>
\begin{pmatrix}
  1 & 1 \\
  1 & 1 
\end{pmatrix}
= [aba] 
\]

For cycle rewriting a similar argument can be given, which is based on
the idea that the trace, i.e., the sum of the diagonal, of a matrix
decreases \cite{sz15}.

A natural question to ask is how such matrices can be obtained. We
will later discuss how SMT solvers can be employed to automatically
generate the required weights.

In the following, we will generalize this method in two ways: we will
replace the natural numbers by an arbitrary semiring -- an observation
that has already been made in the context of string rewriting -- and
we will make the step from string to graph rewriting.

\subsection{Ordered Semirings}

We continue by defining semirings, the algebraic structures in which
we will evaluate the graphs occurring in transformation sequences, and
orders on them.

A (partial) \emph{order} is a reflexive, transitive and antisymmetric
relation.  If $\leq$ is an order, then we denote by $<$ its strict
subrelation e.g. $x<y$ if and only if $x \leq y \land x \neq y$. An order is 
\emph{well-founded} if it does not allow infinite, strictly decreasing 
sequences $x_0 > x_1 > x_2 > \cdots$.

\begin{definition}
  A \emph{semiring} is a tuple $\tpl{S,\oplus,\otimes,0,1}$, where
    $S$ is the (finite or infinite) carrier set, 
    $\tpl{S,\oplus,0}$ is a commutative monoid,
    $\tpl{S,\otimes,1}$ is a monoid,
    $\otimes$ distributes over $\oplus$ and
    $0$ is an annihilator for $\otimes$.
  That is, the following laws hold for all $x,y,z\in S$:
   \begin{align*}
    (x \oplus y) \oplus z &= x \oplus (y \oplus z)  &  0 \oplus x &= x  &  x 
    \otimes 0 &= 0  \\
    (x \otimes y) \otimes z &= x \otimes (y \otimes z)  &  x \oplus 0 &= x  &  
    0 \otimes x &= 0  \\
	(x \oplus y) \otimes z &= (x \otimes z) \oplus (y \otimes z)  &  1 \otimes x 
	&= x  &  x \oplus y &= y \oplus x  \\
	z \otimes (x \oplus y) &= (z \otimes x) \oplus (z \otimes y)  &  x \otimes 1 
	&= x  &&
	\end{align*}
  A semiring $\tpl{S,\oplus,\otimes,0,1}$ is \emph{commutative} if
  $\otimes$ is commutative (that is, if $x \otimes y = y \otimes x$,
  for all $x,y\in S$).
  
  We will often confuse a semiring with its carrier set, that is, $S$
  can refer to both the semiring $\tpl{S,\oplus,\otimes,0,1}$ and the
  carrier set $S$.
\end{definition}

From now onwards we consider only commutative semirings.

In order to come up with termination arguments, we need a partial
order on the semirings that has to be compatible with its operations.

\begin{definition}
  A semiring $\tpl{S,\oplus,\otimes,0,1,\leq}$ is an \emph{ordered
    semiring} if $\tpl{S,\oplus,\otimes,0,1}$ is a semiring and
  ${\leq}\in S\times S$ is a partial order on $S$ such that for all
  $x,y,u,z\in S$:
  \begin{itemize}
  \item $x\leq y$ implies $x \oplus u \leq y \oplus u$ and
    $x \otimes z \leq y \otimes z$ for $z\ge 0$.
  \end{itemize}
  The ordered semiring $S$ is \emph{strongly ordered}, if
  \begin{itemize}
  \item
    $x < y$, $z < u$ implies
    $x \oplus z < y \oplus u$.
  \end{itemize}  
  The ordered semiring $S$ is \emph{strictly ordered}, if in addition
  \begin{itemize}
  \item $x < y$ implies $x \oplus z < y \oplus z$ for all $z\in S$.
  \end{itemize}

  We furthermore define those elements that preserve (strict)
  inequality under multiplication:
  \begin{eqnarray*}
    S_\le & = & \left\{ z\in S\mid \forall x,y\in S\colon \left( x \le
        y \implies
        x\otimes z \le y \otimes z \right) \right\} \\
    S_< & = & \left\{ z\in S\mid \forall x,y\in S\colon \left( x < y
      \implies x\otimes z < y \otimes z \right) \right\}
  \end{eqnarray*}
\end{definition}

In an ordered semiring $S_\le$ contains at least all elements that are
larger or equal than $0$. Furthermore it is easy to see that
$1\in S_<\subseteq S_\le$ and $S_\le$, $S_<$ are both closed under
multiplication.

\begin{example}
  Examples of semirings which play a role in termination proving are:
  \begin{itemize}
  \item The natural numbers form a semiring
    $\tpl{\nats,+,\cdot,0,1,\leq}$, where $\leq$ is the standard
    ordering of the natural numbers. We will call this semiring the
    arithmetic semiring (on the natural numbers).  This is a strictly
    ordered semiring because both $+$ and $\cdot$ are monotone for
    $\le$ and $+$ is monotone for $<$.

    Furthermore $S_\le = \nats$, $S_< = \nats\backslash\{0\}$.
  \item
    The tropical semiring (on the natural numbers) is:
    \[
      T_{\nats}=\tpl{\nats\cup\{\infty\},\mathrm{min},+,\infty,0,\leq}, 
    \]
    where $\leq$ is the usual ordering of the natural numbers.  The
    tropical semiring is not strictly ordered, because, for example,
    $2 < 3$ but $\min(1,2) \not< \min(1,3)$. It is however still
    strongly ordered.

    Furthermore $S_\le = \nats\cup\{\infty\}$, $S_< = \nats$. Note
    that the unit of addition (zero) is $\infty$ and hence $S_<$
    contains strictly more elements than the elements larger or equal
    than zero.
  \item
    The arctic semiring (on the natural numbers) is
    \[
      T_{\nats}=\tpl{\nats\cup\{-\infty\},\mathrm{max},+,-\infty,0,\leq},
    \]
    where $\leq$ is the normal ordering of the natural numbers.
    Like the tropical semiring, the arctic semiring is not strictly
    ordered, but strongly ordered.   

    Furthermore $S_\le = \nats\cup\{-\infty\}$, $S_< = \nats$.
  \end{itemize}
  All semirings above are commutative. We will in the following
  restrict ourselves to commutative semirings, since we are assigning
  weights to graphs by multiplying weights of nodes and edges, and
  nodes and edges are typically unordered.
\end{example}

\section{Weighted Type Graphs}
\label{sec:type-graphs}

Similarly to mapping a word to a matrix, we will associate weights to
graphs, by typing them over a type graph with weights from a semiring.

\begin{definition}
  \label{def:weighted-type-graph}
  Let an ordered semiring $S$ be given.
  A \emph{weighted type graph} $T$ over $S$ is a graph with a weight function 
    $w_T\colon E_T\to S$
  and a designated flower node 
  $\sFlower[T] \in V$, such that 
  for each label $A\in\Lambda$ there exists a designated edge $e_A$ with 
  $\fSrc[T]{e_A} = \sFlower[T]$,
  $\fTgt[T]{e_A} = \sFlower[T]$,
  $\fLab[T]{e_A} = A$ and
  $w_T(e_A) \in S_<$.
  
  For a graph $G$, we denote with $\fFlowerM[T]{G}$ (or just
  $\fFlowerM{G}$ if $T$ is clear from the context) the unique
  morphism from $G$ to $T$ that maps each node $v\in V_G$ of $G$
  to the flower node $\sFlower[T]$ and each edge $e\in E_G$, with
  $\fLab[T]{e}=A$, to $e_A$.  Note that, for a morphism $c\colon
  G\to H$, it is always the case that $\fFlowerM[T]{H} \circ c =
  \fFlowerM[T]{G}$.
\end{definition}

Note that every matrix $A$ of dimension $n$ can be associated with an
(unlabelled) type graph with $n$ nodes, where an edge from node~$i$
to~$j$ is assigned weight $A_{i,j}$ (or does not exist if $A_{i,j} =
0$). Hence our idea of weighted type graphs is strongly related with
the matrices of Section~\ref{sec:matrix-interpretations}.

The node $\sFlower[T]$ is called the flower node, since the loops
attached to it look like a flower. Those loops correspond to the
matrix entries at position $(1,1)$ and similar to those entries they
play a specific role. Note that the flower structure also ensures that
\emph{every} graph can be typed over $T$ (compare with the terminal
object in the category of graphs, which is exactly such a flower).

\smallskip
  
With a bit of notation overloading, we assign a weight to each morphism 
$t\colon D\to T$ with codomain $T$ and arbitrary domain $D$ as follows:
\[
  w_T(t) = \prod_{e \in E_D} w_T(t(e)).
\]
That is, we multiply the weights of all edges in the image of $t$ with respect 
to $\otimes$.

Finally, the weight of a graph $G$ with respect to $T$ is defined by
summing up the weights of all morphisms from $G$ to $T$ with respect to 
$\oplus$:
\[
  w_T(G) = \sum_{t_G\colon G\to T} w_T(t_G).
\]
The subscript $T$ of $w_T$ will be omitted if clear from the context.

\newcommand\ExpRuleABA{
  \begin{tikzpicture}[x=1.2cm,y=-1.2cm,baseline=(a1.south)]
	  \begin{scope}[shift={(-2.5,0)}]
	    \node[gnode] (a1) at (-1,0) {} ;
	    \node[glab,below] (lab1) at (a1.south) {$1$} ;
	    \node[gnode] (a2) at ( 0,0) {} ;
	    \node[gnode] (a3) at ( 1,0) {} ;
	    \node[glab,below] (lab3) at (a3.south) {$2$} ;
	    \draw[gedge] (a1) -- node[glab,above] (labA1) {$a$} (a2) ;
	    \draw[gedge] (a2) -- node[glab,above] (labA2) {$a$} (a3) ;
	    \graphbox[l]{(a1) (lab1) (a2) (a3) (lab3) (labA1) (labA2)}  
	  \end{scope}
	  \begin{scope}
	    \node[gnode] (a1) at (-.4,0) {} ;
	    \node[glab,below] (lab1) at (a1.south) {$1$} ;
	    \node[gnode] (a3) at (.4,0) {} ;
	    \node[glab,below] (lab3) at (a3.south) {$2$} ;
	    \graphbox[i]{(a1) (lab1) (a3) (lab3)}  
	  \end{scope}
	  \begin{scope}[shift={(2.5,0)}]
	    \node[gnode] (a1) at (-1,0) {} ;
	    \node[glab,below] (lab1) at (a1.south) {$1$} ;
	    \node[gnode] (a2) at ( 0,0) {} ;
	    \node[gnode] (a3) at ( 1,0) {} ;
	    \node[gnode] (a4) at ( 2,0) {} ;
	    \node[glab,below] (lab4) at (a4.south) {$2$} ;
	    \draw[gedge] (a1) -- node[glab,above] (labA1) {$a$} (a2) ;
	    \draw[gedge] (a2) -- node[glab,above] (labA2)  {$b$} (a3) ;
	    \draw[gedge] (a3) -- node[glab,above] (labA2)  {$a$} (a4) ;
	    \graphbox[r]{(a1) (lab1) (a2) (a3) (a4) (lab4) (labA1)
              (labA2) (labA2)}          
	  \end{scope}
	  \draw[->] \substy{(i.west)}{0} -- \substy{(l.east)}{0} ;
	  \draw[->] \substy{(i.east)}{0} -- \substy{(r.west)}{0} ;
  \end{tikzpicture}
}


\begin{example}
  \label{ex:weighted-aa-aba-1}
  We give a small example for the weight of a graph, for which we use
  the arithmetic semiring.
\end{example}

  \begin{wrapfigure}{r}{0.30\textwidth}
    \vspace{-.7cm}
    \begin{tabular}{rl}
      $T = {}$ &
        \begin{tikzpicture}[x=1.5cm,y=-1.2cm,baseline=(1.south)]
          \node[gnode] (1) at (0,0) {} ; \node[gnode] (2) at (1.5,0) {}
          ; \draw[gedge] (1) .. controls +(70:1cm) and +(110:1cm) ..
          node[arlab,above] {$a^1$} (1) ; \draw[gedge] (1) .. controls
          +(250:1cm) and +(290:1cm) ..  node[arlab,below] {$b^1$} (1) ;
          \draw[gedge] (2) to[bend left=20] node[arlab,below] {$a^1$}
          (1) ; \draw[gedge] (1) to[bend left=20] node[arlab,above]
          {$a^1$} (2) ;
        \end{tikzpicture}
      \end{tabular}
    \vspace{-.9cm}
  \end{wrapfigure}
  Consider for instance the type graph $T$. Edges are labelled $a,b$
  and the weights, in this case natural numbers, are given as
  superscripts. Consider also the left-hand side $L$ of rule $\rho$
  below, consisting of two $a$-edges (the graph rewriting analogue of
  the string rewriting rule $aa\to aba$ considered in
  Section~\ref{sec:matrix-interpretations}). There are five morphisms
  $L\to T$, each having weight $1$, as they are calculated by
  multiplying the weights of two $a$-edges which also have weight $1$.
  Hence the weight of $L$ with respect to $T$ is $w_T(L) =
  1+1+1+1+1=5$. More details on this are given in Example
  \ref{ex:weighted-aa-aba-2}.
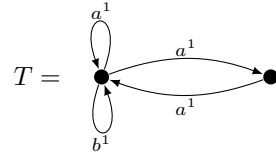
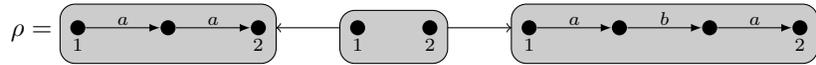
    \begin{align*}
      \rho &= \ExpRuleABA 
    \end{align*}
If we glue two graphs $G_1,G_2$ in order to obtain $G$, the weight of
$G$ can be obtained from the weights of $G_1,G_2$.

\newcommand{\propertiestypegraph}{Let $S$ be an ordered commutative
  semiring and $T$ a weighted type graph over $S$.

\noindent\begin{minipage}{0.6\textwidth}
  \begin{enumerate}[(i)]
  \item Whenever $S$ is strongly ordered, for all graphs $G$,
    $\fFlowerM[T]{G}\colon G\to T$ exists and $w_T(\fFlowerM[T]{G})
    \in S_<$.
  \item \label{it:stable-pushout} Given the following diagram, where
    the square is a pushout and $G_0$ is discrete, it holds that
    $w_T(t) = w_T(t\circ\phi_1) \otimes w_T(t\circ\phi_2)$.
  \end{enumerate}
\end{minipage} \quad
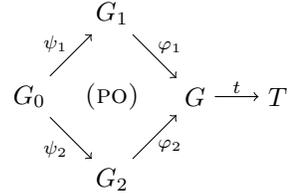
\begin{minipage}{0.4\textwidth}
    \begin{tikzpicture}[x=1.1cm,y=-1.1cm,baseline=(g0)]
      \node (g0) at (0, 0) {$G_0$} ;
      \node (g1) at (1,-1) {$G_1$} ;
      \node (g2) at (1, 1) {$G_2$} ;
      \node (g)  at (2, 0) {$G$}   ;
      \node (t)  at (3, 0) {$T$}   ;
      \draw[->] (g0) -- node[arlab,above left] {$\psi_1$} (g1) ;
      \draw[->] (g0) -- node[arlab,below left] {$\psi_2$} (g2) ;
      \draw[->] (g1) -- node[arlab,above right] {$\phi_1$} (g) ;
      \draw[->] (g2) -- node[arlab,below right] {$\phi_2$} (g) ;
      \draw[->] (g)  -- node[arlab,above] {$t$} (t) ;
      \node at (1,0) { \textsc{(po)} } ;
    \end{tikzpicture}
\end{minipage}
}  
\begin{lemma}[Properties of weighted type graphs]
  \label{lem:properties-type-graph}
  \propertiestypegraph
\end{lemma}
Since property~(\ref{it:stable-pushout}) above only holds if $G_0$ is
discrete we restrict to discrete graphs~$I$ in the rule
interface.\footnote{Compare also with the ``stable under pushouts''
  property of \cite{bkz14}.}

While the process of obtaining the weight of a graph corresponds to
calculating the matrix of a word and summing up all its entries, we
also require a way to be more discriminating, i.e., to access separate
matrix entries. Evaluating a string-like graph would mean to fix its
entry and exit node within the type graph (similarly to fixing two
matrix indices). However, in graph rewriting, we have interfaces of
arbitrary size. Hence, we do not index over pairs of nodes, but over
arbitrary interface graphs, and compute the weight of a graph $L$ with
respect to a typed interface $I$. 

\begin{definition} Let $\phi\colon I\to L$ and $t\colon I\to T$ be graph 
morphisms, where $T$ is a weighted type graph. We define:\\
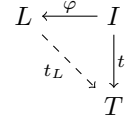
\begin{minipage}[b]{0.77\textwidth}
  \[
  w_t(\phi) = \sum_{\substack{t_L\colon L \to T\\ 
      t_L \circ \phi = t}} w_T(t_L).
  \]
\end{minipage} 
\begin{minipage}[b]{0.2\textwidth}
    \begin{tikzpicture}[x=1.2cm,y=-1.2cm,baseline=(ar.north)]
      \node (g0) at (0, 0) {$L$} ;
      \node (g1) at (1, 0) {$I$} ;
      \node (g2) at (1, 1) {$T$} ;
      \draw[->] (g1) -- node[arlab,above] {$\phi$} (g0) ;
      \draw[->] (g1) -- node[arlab,right] (ar) {$t$} (g2) ;
      \draw[->,dashed] (g0) -- node[arlab,below left] {$t_L$} (g2) ;
    \end{tikzpicture}
\end{minipage}
\end{definition}

Finally, we can define what it means that a rule is decreasing,
analogous to the condition $[\ell] > [r]$ introduced in
Section~\ref{sec:matrix-interpretations}. In addition we also
introduce non-increasingness, a concept that will be needed in the
following for so-called relative termination arguments.

\begin{definition}
  Let a rule $\rho = L\arleft[\phi_L] I \arright[\phi_R] R$, an
  ordered commutative semiring $S$ and a weighted type graph $T$ over
  $S$ be given.
  \begin{enumerate}[(i)]
  \item
    The rule $\rho$ is \emph{non-increasing} if
      for all $t_I\colon I\to T$ it holds that 
        $w_{t_I}(\phi_L) \geq w_{t_I}(\phi_R)$.
  \item
    The rule $\rho$ is \emph{decreasing} if it is 
    non-increasing, and
      $w_{\fFlowerM{I}}(\phi_L) > w_{\fFlowerM{I}}(\phi_R)$.
  \end{enumerate}
\end{definition}

\begin{example}
  \label{ex:weighted-aa-aba-2}
  We come back to Example~\ref{ex:weighted-aa-aba-1} and check whether
  rule $\rho$ is decreasing. For this we have to consider the
  following four morphisms $t\colon I\to T$ from the two-node
  interface into the weighted type graph $T$:
  \begin{itemize}
  \item The flower morphism $\fFlowerM{I}$ which maps both interface
    nodes to the left node of $T$. In this case we have
    $w_{\fFlowerM{I}}(\phi_L) = 2 > 1 = w_{\fFlowerM{I}}(\phi_R)$.
  \item Furthermore there are three other morphisms $t_1,t_2,t_3\colon
    I\to T$ mapping the two interface nodes either both to the right
    node of $T$, or the first interface node to the left and the
    second interface node to the right node of $T$, or vice versa.  In
    all these cases we have $w_{t_i}(\phi_L) = 1 = w_{t_i}(\phi_R)$.
  \end{itemize}
  Hence, the rule is decreasing. Note also that these weights
  correspond exactly to the weights of the multiplied matrices in
  Section~\ref{sec:matrix-interpretations}.
\end{example}

Finally, we have to show that applying a decreasing rule also
decreases the overall weight of a graph. For a non-increasing rule the
weight might also remain the same.

\newcommand{\lemdecreasing}{Let $S$ be a strictly ordered commutative
  semiring and $T$ a weighted type graph over $S$, where all weights
  are contained in $S_\le$.  Furthermore, let $\rho$ be a rule such
  that $G \Rightarrow_\rho H$.
  \begin{enumerate}[(i)]
  \item
    If $\rho$ is non-increasing, then $w_T(G) \geq w_T(H)$.
  \item
    If $\rho$ is decreasing, then $w_T(G) > w_T(H)$.
  \end{enumerate}
}

\begin{lemma}
  \label{lem:decreasing}
  \lemdecreasing
\end{lemma}

From this lemma we can prove our main theorem that is based on the
well-known concept of relative termination \cite{ges90,z:termination}:
if we can find a type graph for which some rules are decreasing and
the rest is non-increasing, we can remove the decreasing rules without
affecting termination. We are then left with a smaller set of rules
for which termination can either be shown with a different type graph
or with some other technique entirely.

\newcommand{\thmterminating}{
  Let $S$ be a strictly ordered commutative semiring with a
  well-founded order $\leq$ and $T$ a weighted type graph over $S$,
  where all weights are contained in $S_\le$.
  Let $R$ be a set of graph transformation rules, partitioned in two
  sets $R^{<}$ and $R^{=}$.  Assume that all rules of $R^{<}$ are
  decreasing and all rules of $R^{=}$ are non-increasing.  Then $R$ is
  terminating if and only if $R^{=}$ is terminating.
}

\begin{theorem}
  \label{thm:terminating}
  \thmterminating
\end{theorem}

A special case of the theorem is when $R^{=}=\emptyset$.  Then the
statement of the theorem is that a graph transformation system $R$ is
terminating if all its rules are decreasing with respect to a strictly
ordered commutative semiring $S$ and type graph $T$ over $S$.

\section{Using Strongly Ordered Semirings}
\label{sec:non-strict}

In the last section the semirings were required to be strictly
ordered.  In this section we consider what happens when we weaken this
requirement and also allow non-strictly ordered semirings, which must
however be strongly ordered. This allows us to work with the tropical
and arctic semiring. It turns out that we obtain similar results to
above if we strengthen the notion of decreasing.

\begin{definition}
  Let a rule $\rho = L\arleft[\phi_L] I \arright[\phi_R] R$, an
  ordered commutative semiring $S$ and a weighted type graph $T$ over
  $S$ be given.  The rule $\rho$ is \emph{strongly decreasing} (with
  respect to $T$) if for all $t_I\colon I\to T$ it holds that
  $w_{t_I}(\phi_L) > w_{t_I}(\phi_R)$.
\end{definition}

Using this new notion of decreasingness we can also formulate a termination
argument, which is basically equivalent to the termination argument
we presented in \cite{bkz14}.

\newcommand{\lemdecreasingnonstrict}{ Let $S$ be a strongly ordered
  commutative semiring and $T$ a weighted type graph over $S$, where
  all weights are contained in $S_<$.  Furthermore, let $\rho$ be a
  rule such that $G \Rightarrow_\rho H$.
  \begin{enumerate}[(i)]
  \item If $\rho$ is non-increasing, then $w_T(G) \geq w_T(H)$.
  \item If $\rho$ is strongly decreasing, then $w_T(G) > w_T(H)$.
  \end{enumerate}
}

\begin{lemma}
  \label{lem:decreasing-nonstrict}
  \lemdecreasingnonstrict
\end{lemma}

Now it is easy to prove a theorem analogous to
Theorem~\ref{thm:terminating}, using
Lemma~\ref{lem:decreasing-nonstrict} instead of
Lemma~\ref{lem:decreasing}. 

\begin{theorem}\label{thm:terminating-weak}
  Let $S$ be a strongly ordered commutative semiring with a
  well-founded order $\leq$ and $T$ a weighted type graph over $S$,
  where all weights are contained in $S_<$.  Let $R$ be a set of graph
  transformation rules, partitioned in two sets $R^{<}$ and $R^{=}$.
  Assume that all rules of $R^{<}$ are strongly decreasing and all
  rules of $R^{=}$ are non-increasing.  Then $R$ is terminating if and
  only if $R^{=}$ is terminating.
\end{theorem}

We have partially recovered the termination analysis from one of our
earlier papers \cite{bkz14}. In order to explain the connection, let
us consider what it means for a rule
$\rho = L\arleft[\phi_L] I \arright[\phi_R] R$ to be non-increasing in
the tropical semiring where $\oplus$ is $\min$ and $\otimes$ is $+$:
for each $t\colon I\to T$ into a weighted type graph $T$ it must hold
that
\[ \min_{\substack{t_L\colon L\to T\\ t_L\circ\phi_L = t}} w_T(t_L)
\ge \min_{\substack{t_R\colon R\to T\\ t_R\circ\phi_R = t}} w_T(t_R)
\] 
where $w_T(t_L)$ is the weight of the morphism $t_L$, obtained by
summing up (via $+$) the weights of all edges in the image of $t_L$.

If the sets are non-empty, a different way of expressing that the
minimum of the first set is larger or equal than the minimum of the
second set, is to say that for each morphism $t_L\colon L\to T$ with
$t_L\circ\phi_L = t$ there exists a morphism $t_R\colon R\to T$ with
$t_R\circ\phi_R = t$ and $w_T(t_L) \ge w_T(t_R)$. This results in the
notion of tropically non-increasing of~\cite{bkz14}.

A proper generalization, taking also empty sets into account, is made
in \cite{eo:generalized-weighted-termination-gts}.

Comparing the results of Theorems~\ref{thm:terminating}
and~\ref{thm:terminating-weak} we notice the following: as underlying
semiring $S$ we can take either a strictly ordered or a strongly
ordered one, but if we choose a strongly ordered semiring, the
termination argument might become more difficult, since for every
morphism from the left-hand side to the type graph there must exist a
compatible, strictly smaller morphism from the right-hand side to the
type graph. In addition, we have to be more restrictive with the
weights allowed in the type graph.

\section{Examples}
\label{sec:example}

We give examples to show that with a weighted type graph over a
strictly ordered semiring (such as the arithmetic semiring), we can
prove termination on some graph transformation systems where strongly
ordered semirings fail. We start with a graph transformation system
for which a termination argument can be found using both variants.
Then we will modify some rules and explain why weighted type graphs
over strongly ordered semirings can not find a termination argument
for the modified system.

\newcommand\ExpRuleOne{
  \begin{tikzpicture}[x=1.2cm,y=-1.2cm,baseline=(a1)]
	  \begin{scope}[shift={(-2.5,0)}]
	    \node[gnode] (a1) at (-1,0) {} ;
	    \node[glab,below] (lab1) at (a1.south) {$1$} ;
	    \node[gnode] (a2) at ( 0,0) {} ;
	    \node[gnode] (a3) at ( 1,0) {} ;
	    \node[glab,below] (lab3) at (a3.south) {$2$} ;
	    \draw[gedge] (a2) -- node[glab,above] (labA1) {$0$} (a1) ;
	    \draw[gedge] (a3) -- node[glab,above] (labA2) {$\mathit{count}$} (a2) ;
	    \draw[gedge] (a2) .. controls +(60:.6cm) and +(120:.6cm) ..  
	    node[glab,above] (labA3) {$\mathit{incr}$} (a2) ;
	    \graphbox[l]{(a1) (lab1) (a2) (a3) (lab3) (labA1) (labA2) (labA3)}  
	  \end{scope}
	  \begin{scope}
	    \node[gnode] (a1) at (-.4,0) {} ;
	    \node[glab,below] (lab1) at (a1.south) {$1$} ;
	    \node[gnode] (a3) at (.4,0) {} ;
	    \node[glab,below] (lab3) at (a3.south) {$2$} ;
	    \graphbox[i]{(a1) (lab1) (a3) (lab3)}  
	  \end{scope}
	  \begin{scope}[shift={(2.5,0)}]
	    \node[gnode] (a1) at (-1,0) {} ;
	    \node[glab,below] (lab1) at (a1.south) {$1$} ;
	    \node[gnode] (a2) at ( 0,0) {} ;
	    \node[gnode] (a3) at ( 1,0) {} ;
	    \node[glab,below] (lab3) at (a3.south) {$2$} ;
	    \draw[gedge] (a2) -- node[glab,above] (labA1) {$1$} (a1) ;
	    \draw[gedge] (a3) -- node[glab,above] (labA2)  {$\mathit{count}$} (a2) ;
	    \graphbox[r]{(a1) (lab1) (a2) (a3) (lab3) (labA1) (labA2)}          
	  \end{scope}
	  \draw[->] \substy{(i.west)}{0} -- \substy{(l.east)}{0} ;
	  \draw[->] \substy{(i.east)}{0} -- \substy{(r.west)}{0} ;
  \end{tikzpicture}
}

\newcommand\ExpRuleTwo{
  \begin{tikzpicture}[x=1.2cm,y=-1.2cm,baseline=(a1)]
	  \begin{scope}[shift={(-2.5,0)}]
	    \node[gnode] (a1) at (-1,0) {} ;
	    \node[glab,below] (lab1) at (a1.south) {$1$} ;
	    \node[gnode] (a2) at ( 0,0) {} ;
	    \node[gnode] (a3) at ( 1,0) {} ;
	    \node[glab,below] (lab3) at (a3.south) {$2$} ;
	    \draw[gedge] (a2) -- node[glab,above] (labA1) {$1$} (a1) ;
	    \draw[gedge] (a3) -- node[glab,above] (labA2) {$\mathit{count}$} (a2) ;
	    \draw[gedge] (a2) .. controls +(60:.6cm) and +(120:.6cm) ..  
	    node[glab,above] (labA3) {$\mathit{incr}$} (a2) ;
	    \graphbox[l]{(a1) (lab1) (a2) (a3) (lab3) (labA1) (labA2) (labA3)}  
	  \end{scope}
	  \begin{scope}
	    \node[gnode] (a1) at (-.4,0) {} ;
	    \node[glab,below] (lab1) at (a1.south) {$1$} ;
	    \node[gnode] (a3) at (.4,0) {} ;
	    \node[glab,below] (lab3) at (a3.south) {$2$} ;
	    \graphbox[i]{(a1) (lab1) (a3) (lab3)}  
	  \end{scope}
	  \begin{scope}[shift={(2.5,0)}]
	    \node[gnode] (a1) at (-1,0) {} ;
	    \node[glab,below] (lab1) at (a1.south) {$1$} ;
	    \node[gnode] (a2) at ( 0,0) {} ;
	    \node[gnode] (a3) at ( 1,0) {} ;
	    \node[glab,below] (lab3) at (a3.south) {$2$} ;
	    \draw[gedge] (a2) -- node[glab,above] (labA1) {$c$} (a1) ;
	    \draw[gedge] (a3) -- node[glab,above] (labA2)  {$\mathit{count}$} (a2) ;
	    \graphbox[r]{(a1) (lab1) (a2) (a3) (lab3) (labA1) (labA2)}          
	  \end{scope}
	  \draw[->] \substy{(i.west)}{0} -- \substy{(l.east)}{0} ;
	  \draw[->] \substy{(i.east)}{0} -- \substy{(r.west)}{0} ;
  \end{tikzpicture}
}

\newcommand\ExpRuleThree{
  \begin{tikzpicture}[x=1.2cm,y=-1.2cm,baseline=(a1.south)]
	  \begin{scope}[shift={(-2.5,0)}]
	    \node[gnode] (a1) at (-1,0) {} ;
	    \node[glab,below] (lab1) at (a1.south) {$1$} ;
	    \node[gnode] (a2) at ( 0,0) {} ;
	    \node[gnode] (a3) at ( 1,0) {} ;
	    \node[glab,below] (lab3) at (a3.south) {$2$} ;
	    \draw[gedge] (a2) -- node[glab,above] (labA1) {$0$} (a1) ;
	    \draw[gedge] (a3) -- node[glab,above] (labA2) {$c$} (a2) ;
	    \graphbox[l]{(a1) (lab1) (a2) (a3) (lab3) (labA1) (labA2)}  
	  \end{scope}
	  \begin{scope}
	    \node[gnode] (a1) at (-.4,0) {} ;
	    \node[glab,below] (lab1) at (a1.south) {$1$} ;
	    \node[gnode] (a3) at (.4,0) {} ;
	    \node[glab,below] (lab3) at (a3.south) {$2$} ;
	    \graphbox[i]{(a1) (lab1) (a3) (lab3)}  
	  \end{scope}
	  \begin{scope}[shift={(2.5,0)}]
	    \node[gnode] (a1) at (-1,0) {} ;
	    \node[glab,below] (lab1) at (a1.south) {$1$} ;
	    \node[gnode] (a2) at ( 0,0) {} ;
	    \node[gnode] (a3) at ( 1,0) {} ;
	    \node[glab,below] (lab3) at (a3.south) {$2$} ;
	    \draw[gedge] (a2) -- node[glab,above] (labA1) {$1$} (a1) ;
	    \draw[gedge] (a3) -- node[glab,above] (labA2)  {$0$} (a2) ;
	    \graphbox[r]{(a1) (lab1) (a2) (a3) (lab3) (labA1) (labA2)}          
	  \end{scope}
	  \draw[->] \substy{(i.west)}{0} -- \substy{(l.east)}{0} ;
	  \draw[->] \substy{(i.east)}{0} -- \substy{(r.west)}{0} ;
  \end{tikzpicture}
}

\newcommand\ExpRuleFour{
  \begin{tikzpicture}[x=1.2cm,y=-1.2cm,baseline=(a1.south)]
	  \begin{scope}[shift={(-2.5,0)}]
	    \node[gnode] (a1) at (-1,0) {} ;
	    \node[glab,below] (lab1) at (a1.south) {$1$} ;
	    \node[gnode] (a2) at ( 0,0) {} ;
	    \node[gnode] (a3) at ( 1,0) {} ;
	    \node[glab,below] (lab3) at (a3.south) {$2$} ;
	    \draw[gedge] (a2) -- node[glab,above] (labA1) {$1$} (a1) ;
	    \draw[gedge] (a3) -- node[glab,above] (labA2) {$c$} (a2) ;
	    \graphbox[l]{(a1) (lab1) (a2) (a3) (lab3) (labA1) (labA2)}  
	  \end{scope}
	  \begin{scope}
	    \node[gnode] (a1) at (-.4,0) {} ;
	    \node[glab,below] (lab1) at (a1.south) {$1$} ;
	    \node[gnode] (a3) at (.4,0) {} ;
	    \node[glab,below] (lab3) at (a3.south) {$2$} ;
	    \graphbox[i]{(a1) (lab1) (a3) (lab3)}  
	  \end{scope}
	  \begin{scope}[shift={(2.5,0)}]
	    \node[gnode] (a1) at (-1,0) {} ;
	    \node[glab,below] (lab1) at (a1.south) {$1$} ;
	    \node[gnode] (a2) at ( 0,0) {} ;
	    \node[gnode] (a3) at ( 1,0) {} ;
	    \node[glab,below] (lab3) at (a3.south) {$2$} ;
	    \draw[gedge] (a2) -- node[glab,above] (labA1) {$c$} (a1) ;
	    \draw[gedge] (a3) -- node[glab,above] (labA2)  {$0$} (a2) ;
	    \graphbox[r]{(a1) (lab1) (a2) (a3) (lab3) (labA1) (labA2)}          
	  \end{scope}
	  \draw[->] \substy{(i.west)}{0} -- \substy{(l.east)}{0} ;
	  \draw[->] \substy{(i.east)}{0} -- \substy{(r.west)}{0} ;
  \end{tikzpicture}
}

\begin{example}\label{ex:addOneOnce}
  As an example we take a system consisting of several counters, which
  represent their current value by a finite number of bits. Each
  counter may possess an \emph{incr} marker, that can be consumed to
  increment the counter by $1$.
\end{example}

\begin{wrapfigure}{r}{0.55\textwidth}
  \vspace{-1.1cm}
  \begin{tabular}{rl}
    $G = {}$ &
    \begin{tikzpicture}[x=1.2cm,y=-1.2cm,baseline=(b.south)]
      \node[gnode] (a2) at (1.4,0) {} ;
      \node[gnode] (a1) at (2.2,0) {} ;
      \node[gnode] (a0) at (3,0) {} ;
      \node[label] (b) at (-.45,-1) {\ldots} ;
      \node[gnode] (b3) at (0.6,-1) {} ;
      \node[gnode] (b2) at (1.4,-1) {} ;
      \node[gnode] (b1) at (2.2,-1) {} ;
      \node[gnode] (b0) at (3,-1) {} ;
      \node[label] (c) at (.35,-2) {\ldots} ;
      \node[gnode] (c2) at (1.4,-2) {} ;
      \node[gnode] (c1) at (2.2,-2) {} ;
      \node[gnode] (c0) at (3,-2) {} ;
      \node[gnode] (d) at (4,-1) {} ;
      \draw[gedge] (a1) -- node[glab,above] {$1$} (a2) ;
      \draw[gedge] (a0) -- node[glab,above] {$0$} (a1) ;
      \draw[gedge] (b3) -- node[glab,above] {$0$} (b) ;
      \draw[gedge] (b2) -- node[glab,above] {$0$} (b3) ;
      \draw[gedge] (b1) -- node[glab,above] {$1$} (b2) ;
      \draw[gedge] (b0) -- node[glab,above] {$1$} (b1) ;
      \draw[gedge] (c2) -- node[glab,above] {$1$} (c) ;
      \draw[gedge] (c1) -- node[glab,above] {$0$} (c2) ;
      \draw[gedge] (c0) -- node[glab,above] {$1$} (c1) ;
      \draw[gedge] (b0) .. controls +(60:.6cm) and +(120:.6cm) ..  
      node[glab,above] {$\mathit{incr}$} (b0) ;
      \draw[gedge] (c0) .. controls +(60:.6cm) and +(120:.6cm) ..  
      node[glab,above] {$\mathit{incr}$} (c0) ;
      \draw[gedge] (d) -- node[glab,midway,below,right] {$\mathit{count}$} 
      (a0) ;
      \draw[gedge] (d) -- node[glab,above] {$\mathit{count}$} (b0) ;
      \draw[gedge] (d) -- node[glab,midway,above,right] {$\ \mathit{count}$} 
      (c0) ;
    \end{tikzpicture}
  \end{tabular}
  \vspace{-.9cm}
\end{wrapfigure}
One possible graph describing a state of such a system is given
by~$G$. This is just one possible initial graph, since we really
show uniform termination, i.e., termination on all initial graphs,
even those that do not conform to the schema indicated by $G$.

We consider the graph transformation system $\{\rho_1, \rho_2,
\rho_3, \rho_4\}$, adapted from \cite{sz15}, consisting of the
following four rules:
\begin{align*}
  \rho_1 &= \ExpRuleOne 
  \displaybreak[0] \\
  \rho_2 &= \ExpRuleTwo 
  \displaybreak[0] \\
  \rho_3 &= \ExpRuleThree 
  \displaybreak[0] \\
  \rho_4 &= \ExpRuleFour
\end{align*}
  
Each counter may increment at most once. Rules $\rho_1$ and $\rho_2$
specify that a counter (represented by a \emph{count}-labelled
edge) may increment its least significant bit by $1$ if an \emph{incr}
marker was not consumed yet. If the least significant bit is $1$, the
bit is marked by a label \emph{c}, to remember that a carry bit has
to be passed to the following bit.  Rule $\rho_3$ increments the next
bit of the counter by $1$ (if it was $0$ before), while rule
$\rho_4$ shifts the carry bit marker over the next $1$.

\begin{wrapfigure}{r}{4.1cm}
  \vspace{-0.7cm}
  \begin{tabular}{rl}
    $T_{\mathit{trop}} = {}$ &
    \begin{tikzpicture}[x=1.5cm,y=-1.2cm,baseline=(1.south)]
      \node[gnode] (1) at (0,0) {} ;
      \draw[gedge] (1) .. controls +(-20:1cm) and +(20:1cm) .. 
      node[arlab,right] {$\mathit{count}^0$} (1) ;
      \draw[gedge] (1) .. controls +(52:1cm) and +(92:1cm) .. 
      node[arlab,above right] {$\mathit{incr}^2$} (1) ;
      \draw[gedge] (1) .. controls +(124:1cm) and +(164:1cm) .. 
      node[arlab,above left] {$0^0$} (1) ;
      \draw[gedge] (1) .. controls +(196:1cm) and +(236:1cm) .. 
      node[arlab,below left] {$c^2$} (1) ;
      \draw[gedge] (1) .. controls +(268:1cm) and +(308:1cm) ..
      node[arlab,below right] {$1^1$} (1) ;
    \end{tikzpicture}
  \end{tabular}
  \vspace{-.5cm}
\end{wrapfigure}

\noindent The fact that this graph transformation system is
uniformly terminating can be shown using a weighted type graph over
either a strictly or strongly ordered semiring. For example, using a
non-relative termination argument, we evaluate the rules with
respect to the weighted type graph $T_{\mathit{trop}}$ over the tropical
semiring. 

A relative termination argument is even easier: the rules $\rho_1$
and $\rho_2$ can be removed due to the decreasing number of
\emph{incr}-labelled edges. Then we can remove $\rho_3$ due to the
decreasing number of \emph{c}-labelled edges (which remain
constant in $\rho_4$) and afterwards remove $\rho_4$ since it
decreases $1$-labelled edges. With all rules removed, the graph
transformation system has been shown to terminate uniformly.
  
\begin{wrapfigure}{r}{3.8cm}
  \vspace{-1.3cm}
  \begin{tabular}{rl}
    $T_{\mathit{arit}} = {}$ &
      \begin{tikzpicture}[x=1.5cm,y=-1.2cm,baseline=(1.south)]
        \node[gnode] (1) at (0,0) {} ;
        \draw[gedge] (1) .. controls +(-20:1cm) and +(20:1cm) .. 
                     node[arlab,right] {$\mathit{count}^1$} (1) ;
        \draw[gedge] (1) .. controls +(52:1cm) and +(92:1cm) .. 
                     node[arlab,above right] {$\mathit{incr}^3$} (1) ;
        \draw[gedge] (1) .. controls +(124:1cm) and +(164:1cm) .. 
                     node[arlab,above left] {$0^1$} (1) ;
        \draw[gedge] (1) .. controls +(196:1cm) and +(236:1cm) .. 
                     node[arlab,below left] {$c^3$} (1) ;
        \draw[gedge] (1) .. controls +(268:1cm) and +(308:1cm) ..
                     node[arlab,below right] {$1^2$} (1) ;
      \end{tikzpicture}
  \end{tabular}
  \vspace{-.9cm}
\end{wrapfigure}
  
\noindent We now consider the arithmetic semiring and again use a
non-relative termination argument: we evaluate the rules with respect
to the weighted type graph $T_{\mathit{arit}}$, where all weights are just
increased by one with respect to $T_{\mathit{trop}}$. That is due to the fact,
that we are working in the arithmetic semiring and hence have to make
sure that all weights of flower edges are strictly larger than $0$.

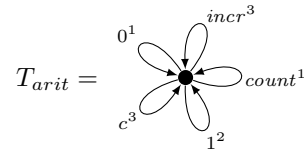
\newcommand\ExpRuleOneAlternated{
  \begin{tikzpicture}[x=1.2cm,y=-1.2cm,baseline=(a1.south)]
	  \begin{scope}[shift={(-2.5,0)}]
	    \node[gnode] (a1) at (-1,0) {} ;
	    \node[glab,below] (lab1) at (a1.south) {$1$} ;
	    \node[gnode] (a2) at ( 0,0) {} ;
	    \node[gnode] (a3) at ( 1,0) {} ;
	    \node[glab,below] (lab3) at (a3.south) {$2$} ;
	    \draw[gedge] (a2) -- node[glab,above] (labA1) {$0$} (a1) ;
	    \draw[gedge] (a3) -- node[glab,above] (labA2) {$\mathit{count}$} (a2) ;
	    \graphbox[l]{(a1) (lab1) (a2) (a3) (lab3) (labA1) (labA2)}  
	  \end{scope}
	  \begin{scope}
	    \node[gnode] (a1) at (-.4,0) {} ;
	    \node[glab,below] (lab1) at (a1.south) {$1$} ;
	    \node[gnode] (a3) at (.4,0) {} ;
	    \node[glab,below] (lab3) at (a3.south) {$2$} ;
	    \graphbox[i]{(a1) (lab1) (a3) (lab3)}  
	  \end{scope}
	  \begin{scope}[shift={(2.5,0)}]
	    \node[gnode] (a1) at (-1,0) {} ;
	    \node[glab,below] (lab1) at (a1.south) {$1$} ;
	    \node[gnode] (a2) at ( 0,0) {} ;
	    \node[gnode] (a3) at ( 1,0) {} ;
	    \node[glab,below] (lab3) at (a3.south) {$2$} ;
	    \draw[gedge] (a2) -- node[glab,above] (labA1) {$1$} (a1) ;
	    \draw[gedge] (a3) -- node[glab,above] (labA2)  {$\mathit{count}$} (a2) ;
	    \graphbox[r]{(a1) (lab1) (a2) (a3) (lab3) (labA1) (labA2)}          
	  \end{scope}
	  \draw[->] \substy{(i.west)}{0} -- \substy{(l.east)}{0} ;
	  \draw[->] \substy{(i.east)}{0} -- \substy{(r.west)}{0} ;
  \end{tikzpicture}
}

\newcommand\ExpRuleTwoAlternated{
  \begin{tikzpicture}[x=1.2cm,y=-1.2cm,baseline=(a1.south)]
	  \begin{scope}[shift={(-2.5,0)}]
	    \node[gnode] (a1) at (-1,0) {} ;
	    \node[glab,below] (lab1) at (a1.south) {$1$} ;
	    \node[gnode] (a2) at ( 0,0) {} ;
	    \node[gnode] (a3) at ( 1,0) {} ;
	    \node[glab,below] (lab3) at (a3.south) {$2$} ;
	    \draw[gedge] (a2) -- node[glab,above] (labA1) {$1$} (a1) ;
	    \draw[gedge] (a3) -- node[glab,above] (labA2) {$\mathit{count}$} (a2) ;
	    \graphbox[l]{(a1) (lab1) (a2) (a3) (lab3) (labA1) (labA2)}  
	  \end{scope}
	  \begin{scope}
	    \node[gnode] (a1) at (-.4,0) {} ;
	    \node[glab,below] (lab1) at (a1.south) {$1$} ;
	    \node[gnode] (a3) at (.4,0) {} ;
	    \node[glab,below] (lab3) at (a3.south) {$2$} ;
	    \graphbox[i]{(a1) (lab1) (a3) (lab3)}  
	  \end{scope}
	  \begin{scope}[shift={(2.5,0)}]
	    \node[gnode] (a1) at (-1,0) {} ;
	    \node[glab,below] (lab1) at (a1.south) {$1$} ;
	    \node[gnode] (a2) at ( 0,0) {} ;
	    \node[gnode] (a3) at ( 1,0) {} ;
	    \node[glab,below] (lab3) at (a3.south) {$2$} ;
	    \draw[gedge] (a2) -- node[glab,above] (labA1) {$c$} (a1) ;
	    \draw[gedge] (a3) -- node[glab,above] (labA2)  {$\mathit{count}$} (a2) ;
	    \graphbox[r]{(a1) (lab1) (a2) (a3) (lab3) (labA1) (labA2)}          
	  \end{scope}
	  \draw[->] \substy{(i.west)}{0} -- \substy{(l.east)}{0} ;
	  \draw[->] \substy{(i.east)}{0} -- \substy{(r.west)}{0} ;
  \end{tikzpicture}
}

\begin{example}\label{ex:addOneInf}
  We will now modify rules $\rho_1$ and $\rho_2$ in order to give an
  example where weighted type graphs over tropical and arctic
  semirings fail to find a termination argument.

  Consider the graph transformation system $\{\rho_{1}',
  \rho_{2}', \rho_{3}', \rho_{4}'\}$ consisting of rules $\rho_3$ and
  $\rho_4$ from Example \ref{ex:addOneOnce} with two additional new
  rules:
  \begin{align*}
    \rho_{1}' &= \ExpRuleOneAlternated &
    \displaybreak[0] \\
    \rho_{2}' &= \ExpRuleTwoAlternated &
    \displaybreak[0] \\
    \rho_{3}' &= \ExpRuleThree &(= \rho_3)
    \displaybreak[0] \\
    \rho_{4}' &= \ExpRuleFour &(= \rho_4)
  \end{align*}
  With respect to Example~\ref{ex:addOneOnce}, the counter may
  increment its value not only once but several times, until the least
  significant bit is permanently marked by the carrier bit label $c$.
  This will eventually happen, since counters are never extended by
  additional digits and carry bits finally accumulate and can not be
  processed.

  We now give a relative termination argument, to show uniform
  termination of this graph transformation system. The termination of
  this system is not obvious as the numbers of the labels $c$, $0$ and
  $1$ increase and decrease depending on the rules used for the
  derivation.
\end{example}

\begin{wrapfigure}{r}{0.45\textwidth}
  \vspace{-1.1cm}
  \begin{tabular}{rl}
    $T' = {}$ &
    \begin{tikzpicture}[x=1.5cm,y=-1.2cm,baseline=(1.south)]
      \node[gnode] (1) at (0,0) {} ;
      \node[gnode] (2) at (1.5,0) {} ;
      \draw[gedge] (1) .. controls +(60:1cm) and +(100:1cm) .. 
                   node[arlab,above] {$\mathit{count}^1$} (1) ;
      \draw[gedge] (1) .. controls +(120:1cm) and +(160:1cm) .. 
                   node[arlab,above left] {$0^1$} (1) ;
      \draw[gedge] (1) .. controls +(180:1cm) and +(220:1cm) .. 
                   node[arlab,left] {$c^1$} (1) ;
      \draw[gedge] (1) .. controls +(240:1cm) and +(280:1cm) .. 
                   node[arlab,below] {$1^1$} (1) ;
      \draw[gedge] (2) to[bend left=10] node[arlab,below] {$1^1$} (1) ;
      \draw[gedge] (2) to[bend left=45] node[arlab,below] {$0^2$} (1) ;
      \draw[gedge] (1) to[bend left=20] node[arlab,above] {$\mathit{count}^1$} 
      (2) ;
      \draw[gedge] (2) .. controls +(40:1cm) and +(80:1cm) .. 
                   node[arlab,above right] {$0^2$} (2) ;
      \draw[gedge] (2) .. controls +(340:1cm) and +(20:1cm) .. 
                   node[arlab,right] {$c^2$} (2) ;
      \draw[gedge] (2) .. controls +(280:1cm) and +(320:1cm) .. 
                   node[arlab,below right] {$1^2$} (2) ;
    \end{tikzpicture}
  \end{tabular}
  \vspace{-.9cm}
\end{wrapfigure}

\noindent First, we evaluate the rules with respect to the following
weighted type graph $T'$ over the arithmetic semiring. Consider for
instance rule $\rho'_1$ and the following four interface morphisms:
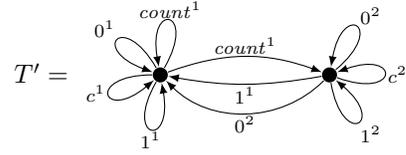
\begin{itemize}
\item $t_0 = \fFlowerM{I}\colon I\to T'$ is the flower morphisms and maps both
  interface node to the left node of $T'$. In this situation we have
  $w_{t_0}(\phi_L) = 1\cdot 1 + 1\cdot 2 = 3 > 2 = 1\cdot 1 + 1\cdot 1
  = w_{t_0}(\phi_R)$ (there are two ways to map the left-hand side in
  such a way that both interface nodes are mapped to the left node,
  resulting in weight~$3$; similar for the right-hand
  side, where we obtain weight~$2$).
\item $t_1\colon I\to T'$ is the morphism that maps the first
  interface node to the right node of $T'$ and the second interface
  node to the left node of $T'$. In this case we have $w_{t_1}(\phi_L)
  = 1\cdot 2 = 2\ge 2 = 1\cdot 2 = w_{t_1}(\phi_R)$.
\item $t_2\colon I\to T'$ is the morphism that maps the first
  interface node to the left node of $T'$ and the second interface
  node to the right node of $T'$. In this case we have
  $w_{t_2}(\phi_L) = 0 \ge 0 = w_{t_2}(\phi_R)$, since there are no
  possibilities to map either the left-hand or the right-hand side.
\item $t_3\colon I\to T'$ is the morphisms that maps both interface
  node to the right node of $T'$. Here we have $w_{t_3}(\phi_L) = 0
  \ge 0 = w_{t_3}(\phi_R)$ (again, there are no fitting matches of the
  left-hand and right-hand side).
\end{itemize}
Hence $\rho'_1$ is decreasing. Similarly we can prove that $\rho'_2$
is decreasing and $\rho'_3,\rho'_4$ are non-increasing, which means
that $\rho'_1,\rho'_2$ can be removed. To show termination of the remaining 
rules $\rho_{3}',\rho_{4}'$ we
can simply use the weighted type graph $T_{\mathit{arit}}$ from Example
\ref{ex:addOneOnce} again.

\smallskip

We found a relative termination argument for
Example~\ref{ex:addOneInf} using a weighted type graph over the
arithmetic semiring. However, there is no way to obtain a termination argument 
with a weighted type graph over either tropical or arctic semirings: in these
cases the weight of any graph is linear in the size of the graph
(since we use only addition and minimum/maximum to determine the
weight of a graph). If we have an interpretation where at least one
rule is decreasing, and the other rules are non-increasing, then in
any derivation, the number of applications of the decreasing rules is
at most linear in the size of the initial graph. However, if we start
with a counter which consists of $n$ bits (all set to $0$), we obtain
a derivation in which $\emph{all}$ of the rules are applied at least
$2^n$ times. 

This means that it is principally impossible to find a proof with
weighted type graphs over the tropical or arctic semiring, even using
relative termination.

\smallskip

The last two examples were inspired by string rewriting and the
example rules could easily be encoded into a string grammar. We give
another final example and prove termination using a weighted type
graph over the arithmetic semiring. We now switch from strings to
trees, staying with a scenario where reductions of exponential length
are possible. In addition we discard the $\mathit{count}$-label as
each counter will be represented by a node with no incoming edge and
we will exploit the dangling edge condition.

\begin{example}\label{ex:counterTree}
  In the next example we interweave our counters into a single treelike 
  structure. Each path from a root node to a leaf can be interpreted as a 
  counter.
\end{example}

  \begin{wrapfigure}{r}{0.45\textwidth}
    \vspace{-.75cm}
    \begin{tabular}{rl}
      $\widehat{G} = {}$ &
      \begin{tikzpicture}[x=1.2cm,y=-1.2cm,baseline=(a.south)]
        \node[gnode] (a) at (4,0) {};
        \node[gnode] (a2) at (3.2,0) {};
        \node[gnode] (b1) at (2.4,-0.6) {};
        \node[gnode] (b2) at (2.4,0.6) {};
        \node[gnode] (c1) at (1.6,-1.0) {};
        \node[gnode] (c2) at (1.6,-0.2) {};
        \node[gnode] (d1) at (1.6,0.2) {};
        \node[label] (d2) at (1.4,1.0) {\ldots};
        \node[label] (e1) at (0.6,0) {\ldots};
        \node[gnode] (e2) at (0.8,-0.55) {};
        \draw[gedge] (a) -- node[glab,above] {$0$} (a2) ;
        \draw[gedge] (a2) -- node[glab,above] {$0$} (b1) ;
        \draw[gedge] (a2) -- node[glab,below] {$0$} (b2) ;
        \draw[gedge] (b2) -- node[glab,above] {$0$} (d1) ;
        \draw[gedge] (b2) -- node[glab,below] {$1$} (d2) ;
        \draw[gedge] (b1) -- node[glab,above] {$1$} (c1) ;
        \draw[gedge] (b1) -- node[glab,below] {$1$} (c2) ;
        \draw[gedge] (c2) -- node[glab,below] {$1$} (e1) ;
        \draw[gedge] (c2) -- node[glab,above] {$0$} (e2) ;
      \end{tikzpicture}
    \end{tabular}
    \vspace{-0.8cm}
  \end{wrapfigure}
  One possible graph describing a state of the modified system is
  given by~$\widehat{G}$. Each counter shares a number of bits with
  other counters, where the least significant bit is shared by all
  counters. Again this is just one possible initial graph, since we
  prove uniform termination.
  
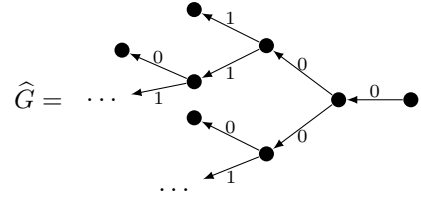
\newcommand\ExpRuleInitTreeOne{
  \begin{tikzpicture}[x=0.8cm,y=-1.2cm,baseline=(a1.south)]
	  \begin{scope}[shift={(-2.3,0)}]
	    \node[gnode] (a1) at (-1,0) {} ;
	    \node[glab,below] (lab1) at (a1.south) {$1$} ;
	    \node[gnode] (a2) at ( 0,0) {} ;
	    \draw[gedge] (a2) -- node[glab,above] (labA1) {$0$} (a1) ;
	    \graphbox[l]{(a1) (lab1) (a2) (labA1) }  
	  \end{scope}
	  \begin{scope}
	    \node[gnode] (a1) at (-.4,0) {} ;
	    \node[glab,below] (lab1) at (a1.south) {$1$} ;
	    \graphbox[i]{(a1) (lab1)}  
	  \end{scope}
	  \begin{scope}[shift={(2.4,0)}]
	    \node[gnode] (a1) at (-1,0) {} ;
	    \node[glab,below] (lab1) at (a1.south) {$1$} ;
	    \node[gnode] (a2) at ( 0,0) {} ;
	    \draw[gedge] (a2) -- node[glab,above] (labA1) {$1$} (a1) ;
	    \graphbox[r]{(a1) (lab1) (a2) (labA1)}          
	  \end{scope}
	  \draw[->] \substy{(i.west)}{0} -- \substy{(l.east)}{0} ;
	  \draw[->] \substy{(i.east)}{0} -- \substy{(r.west)}{0} ;
  \end{tikzpicture}
}

\newcommand\ExpRuleInitTreeTwo{
  \begin{tikzpicture}[x=0.8cm,y=-1.2cm,baseline=(a1.south)]
	  \begin{scope}[shift={(-2.3,0)}]
	    \node[gnode] (a1) at (-1,0) {} ;
	    \node[glab,below] (lab1) at (a1.south) {$1$} ;
	    \node[gnode] (a2) at ( 0,0) {} ;
	    \draw[gedge] (a2) -- node[glab,above] (labA1) {$1$} (a1) ;
	    \graphbox[l]{(a1) (lab1) (a2) (labA1) }  
	  \end{scope}
	  \begin{scope}
	    \node[gnode] (a1) at (-.4,0) {} ;
	    \node[glab,below] (lab1) at (a1.south) {$1$} ;
	    \graphbox[i]{(a1) (lab1)}  
	  \end{scope}
	  \begin{scope}[shift={(2.4,0)}]
	    \node[gnode] (a1) at (-1,0) {} ;
	    \node[glab,below] (lab1) at (a1.south) {$1$} ;
	    \node[gnode] (a2) at ( 0,0) {} ;
	    \draw[gedge] (a2) -- node[glab,above] (labA1) {$c$} (a1) ;
	    \graphbox[r]{(a1) (lab1) (a2) (labA1)}          
	  \end{scope}
	  \draw[->] \substy{(i.west)}{0} -- \substy{(l.east)}{0} ;
	  \draw[->] \substy{(i.east)}{0} -- \substy{(r.west)}{0} ;
  \end{tikzpicture}
}

\newcommand\ExpRuleOneTree{
  \begin{tikzpicture}[x=0.65cm,y=-1.2cm,baseline=(a2.south)]
	  \begin{scope}[shift={(-2.5,0)}]
	    \node[gnode] (a1) at (-1,-0.3) {} ;
	    \node[gnode] (a4) at (-1,0.3) {} ;
	    \node[glab,below] (lab1) at (a1.south) {$1$} ;
	    \node[glab,below] (lab2) at (a4.south) {$2$} ;
	    \node[gnode] (a2) at ( 0,0) {} ;
	    \node[gnode] (a3) at ( 1,0) {} ;
	    \node[glab,below] (lab3) at (a3.south) {$3$} ;
	    \draw[gedge] (a2) -- node[glab,above right] (labA1) {$0$} (a1) ;
	    \draw[gedge] (a2) -- node[glab,below right] (labA3) {$0$} (a4) ;
	    \draw[gedge] (a3) -- node[glab,above] (labA2) {$c$} (a2) ;
	    \graphbox[l]{(a1) (lab1) (a2) (a3) (a4) (lab2) (lab3) (labA1) (labA2) 
	    (labA3)}  
	  \end{scope}
	  \begin{scope}
	    \node[gnode] (a1) at (-.4,-0.3) {} ;
	    \node[glab,below] (lab1) at (a1.south) {$1$} ;
	    \node[gnode] (a2) at (-.4,0.3) {} ;
	    \node[glab,below] (lab2) at (a2.south) {$2$} ;
	    \node[gnode] (a3) at (.4,0) {} ;
	    \node[glab,below] (lab3) at (a3.south) {$3$} ;
	    \graphbox[i]{(a1) (lab1) (a2) (lab2) (a3) (lab3)}  
	  \end{scope}
	  \begin{scope}[shift={(2.5,0)}]
	    \node[gnode] (a1) at (-1,-0.3) {} ;
	    \node[gnode] (a4) at (-1,0.3) {} ;
	    \node[glab,below] (lab1) at (a1.south) {$1$} ;
	    \node[glab,below] (lab2) at (a4.south) {$2$} ;
	    \node[gnode] (a2) at ( 0,0) {} ;
	    \node[gnode] (a3) at ( 1,0) {} ;
	    \node[glab,below] (lab3) at (a3.south) {$3$} ;
	    \draw[gedge] (a2) -- node[glab,above right] (labA1) {$1$} (a1) ;
	    \draw[gedge] (a2) -- node[glab,below right] (labA3) {$0$} (a4) ;
	    \draw[gedge] (a3) -- node[glab,above] (labA2) {$0$} (a2) ;
	    \graphbox[r]{(a1) (lab1) (a2) (a3) (a4) (lab2) (lab3) (labA1) (labA2) 
	    (labA3)}  
	  \end{scope}
	  \draw[->] \substy{(i.west)}{0} -- \substy{(l.east)}{0} ;
	  \draw[->] \substy{(i.east)}{0} -- \substy{(r.west)}{0} ;
  \end{tikzpicture}
}

\newcommand\ExpRuleTwoTree{
  \begin{tikzpicture}[x=0.65cm,y=-1.2cm,baseline=(a2.south)]
	  \begin{scope}[shift={(-2.5,0)}]
	    \node[gnode] (a1) at (-1,-0.3) {} ;
	    \node[gnode] (a4) at (-1,0.3) {} ;
	    \node[glab,below] (lab1) at (a1.south) {$1$} ;
	    \node[glab,below] (lab2) at (a4.south) {$2$} ;
	    \node[gnode] (a2) at ( 0,0) {} ;
	    \node[gnode] (a3) at ( 1,0) {} ;
	    \node[glab,below] (lab3) at (a3.south) {$3$} ;
	    \draw[gedge] (a2) -- node[glab,above right] (labA1) {$0$} (a1) ;
	    \draw[gedge] (a2) -- node[glab,below right] (labA3) {$1$} (a4) ;
	    \draw[gedge] (a3) -- node[glab,above] (labA2) {$c$} (a2) ;
	    \graphbox[l]{(a1) (lab1) (a2) (a3) (a4) (lab2) (lab3) (labA1) (labA2) 
	    (labA3)}  
	  \end{scope}
	  \begin{scope}
	    \node[gnode] (a1) at (-.4,-0.3) {} ;
	    \node[glab,below] (lab1) at (a1.south) {$1$} ;
	    \node[gnode] (a2) at (-.4,0.3) {} ;
	    \node[glab,below] (lab2) at (a2.south) {$2$} ;
	    \node[gnode] (a3) at (.4,0) {} ;
	    \node[glab,below] (lab3) at (a3.south) {$3$} ;
	    \graphbox[i]{(a1) (lab1) (a2) (lab2) (a3) (lab3)}  
	  \end{scope}
	  \begin{scope}[shift={(2.5,0)}]
	    \node[gnode] (a1) at (-1,-0.3) {} ;
	    \node[gnode] (a4) at (-1,0.3) {} ;
	    \node[glab,below] (lab1) at (a1.south) {$1$} ;
	    \node[glab,below] (lab2) at (a4.south) {$2$} ;
	    \node[gnode] (a2) at ( 0,0) {} ;
	    \node[gnode] (a3) at ( 1,0) {} ;
	    \node[glab,below] (lab3) at (a3.south) {$3$} ;
	    \draw[gedge] (a2) -- node[glab,above right] (labA1) {$1$} (a1) ;
	    \draw[gedge] (a2) -- node[glab,below right] (labA3) {$1$} (a4) ;
	    \draw[gedge] (a3) -- node[glab,above] (labA2) {$0$} (a2) ;
	    \graphbox[r]{(a1) (lab1) (a2) (a3) (a4) (lab2) (lab3) (labA1) (labA2) 
	    (labA3)}  
	  \end{scope}
	  \draw[->] \substy{(i.west)}{0} -- \substy{(l.east)}{0} ;
	  \draw[->] \substy{(i.east)}{0} -- \substy{(r.west)}{0} ;
  \end{tikzpicture}
}

\newcommand\ExpRuleThreeTree{
  \begin{tikzpicture}[x=0.65cm,y=-1.2cm,baseline=(a2.south)]
	  \begin{scope}[shift={(-2.5,0)}]
	    \node[gnode] (a1) at (-1,-0.3) {} ;
	    \node[gnode] (a4) at (-1,0.3) {} ;
	    \node[glab,below] (lab1) at (a1.south) {$1$} ;
	    \node[glab,below] (lab2) at (a4.south) {$2$} ;
	    \node[gnode] (a2) at ( 0,0) {} ;
	    \node[gnode] (a3) at ( 1,0) {} ;
	    \node[glab,below] (lab3) at (a3.south) {$3$} ;
	    \draw[gedge] (a2) -- node[glab,above right] (labA1) {$1$} (a1) ;
	    \draw[gedge] (a2) -- node[glab,below right] (labA3) {$0$} (a4) ;
	    \draw[gedge] (a3) -- node[glab,above] (labA2) {$c$} (a2) ;
	    \graphbox[l]{(a1) (lab1) (a2) (a3) (a4) (lab2) (lab3) (labA1) (labA2) 
	    (labA3)}  
	  \end{scope}
	  \begin{scope}
	    \node[gnode] (a1) at (-.4,-0.3) {} ;
	    \node[glab,below] (lab1) at (a1.south) {$1$} ;
	    \node[gnode] (a2) at (-.4,0.3) {} ;
	    \node[glab,below] (lab2) at (a2.south) {$2$} ;
	    \node[gnode] (a3) at (.4,0) {} ;
	    \node[glab,below] (lab3) at (a3.south) {$3$} ;
	    \graphbox[i]{(a1) (lab1) (a2) (lab2) (a3) (lab3)}  
	  \end{scope}
	  \begin{scope}[shift={(2.5,0)}]
	    \node[gnode] (a1) at (-1,-0.3) {} ;
	    \node[gnode] (a4) at (-1,0.3) {} ;
	    \node[glab,below] (lab1) at (a1.south) {$1$} ;
	    \node[glab,below] (lab2) at (a4.south) {$2$} ;
	    \node[gnode] (a2) at ( 0,0) {} ;
	    \node[gnode] (a3) at ( 1,0) {} ;
	    \node[glab,below] (lab3) at (a3.south) {$3$} ;
	    \draw[gedge] (a2) -- node[glab,above right] (labA1) {$c$} (a1) ;
	    \draw[gedge] (a2) -- node[glab,below right] (labA3) {$0$} (a4) ;
	    \draw[gedge] (a3) -- node[glab,above] (labA2) {$0$} (a2) ;
	    \graphbox[r]{(a1) (lab1) (a2) (a3) (a4) (lab2) (lab3) (labA1) (labA2) 
	    (labA3)}  
	  \end{scope}
	  \draw[->] \substy{(i.west)}{0} -- \substy{(l.east)}{0} ;
	  \draw[->] \substy{(i.east)}{0} -- \substy{(r.west)}{0} ;
  \end{tikzpicture}
}

\newcommand\ExpRuleFourTree{
  \begin{tikzpicture}[x=0.65cm,y=-1.2cm,baseline=(a2.south)]
	  \begin{scope}[shift={(-2.5,0)}]
	    \node[gnode] (a1) at (-1,-0.3) {} ;
	    \node[gnode] (a4) at (-1,0.3) {} ;
	    \node[glab,below] (lab1) at (a1.south) {$1$} ;
	    \node[glab,below] (lab2) at (a4.south) {$2$} ;
	    \node[gnode] (a2) at ( 0,0) {} ;
	    \node[gnode] (a3) at ( 1,0) {} ;
	    \node[glab,below] (lab3) at (a3.south) {$3$} ;
	    \draw[gedge] (a2) -- node[glab,above right] (labA1) {$1$} (a1) ;
	    \draw[gedge] (a2) -- node[glab,below right] (labA3) {$1$} (a4) ;
	    \draw[gedge] (a3) -- node[glab,above] (labA2) {$c$} (a2) ;
	    \graphbox[l]{(a1) (lab1) (a2) (a3) (a4) (lab2) (lab3) (labA1) (labA2) 
	    (labA3)}  
	  \end{scope}
	  \begin{scope}
	    \node[gnode] (a1) at (-.4,-0.3) {} ;
	    \node[glab,below] (lab1) at (a1.south) {$1$} ;
	    \node[gnode] (a2) at (-.4,0.3) {} ;
	    \node[glab,below] (lab2) at (a2.south) {$2$} ;
	    \node[gnode] (a3) at (.4,0) {} ;
	    \node[glab,below] (lab3) at (a3.south) {$3$} ;
	    \graphbox[i]{(a1) (lab1) (a2) (lab2) (a3) (lab3)}  
	  \end{scope}
	  \begin{scope}[shift={(2.5,0)}]
	    \node[gnode] (a1) at (-1,-0.3) {} ;
	    \node[gnode] (a4) at (-1,0.3) {} ;
	    \node[glab,below] (lab1) at (a1.south) {$1$} ;
	    \node[glab,below] (lab2) at (a4.south) {$2$} ;
	    \node[gnode] (a2) at ( 0,0) {} ;
	    \node[gnode] (a3) at ( 1,0) {} ;
	    \node[glab,below] (lab3) at (a3.south) {$3$} ;
	    \draw[gedge] (a2) -- node[glab,above right] (labA1) {$c$} (a1) ;
	    \draw[gedge] (a2) -- node[glab,below right] (labA3) {$1$} (a4) ;
	    \draw[gedge] (a3) -- node[glab,above] (labA2) {$0$} (a2) ;
	    \graphbox[r]{(a1) (lab1) (a2) (a3) (a4) (lab2) (lab3) (labA1) (labA2) 
	    (labA3)}  
	  \end{scope}
	  \draw[->] \substy{(i.west)}{0} -- \substy{(l.east)}{0} ;
	  \draw[->] \substy{(i.east)}{0} -- \substy{(r.west)}{0} ;
  \end{tikzpicture}
}

  \noindent Let the following graph transformation system 
  $\{\widehat{\rho_{1}}, 
  \widehat{\rho_{2}}, \widehat{\rho_{3}}, \widehat{\rho_{4}}, 
  \widehat{\rho_{5}}, \widehat{\rho_{6}}\}$ be given:
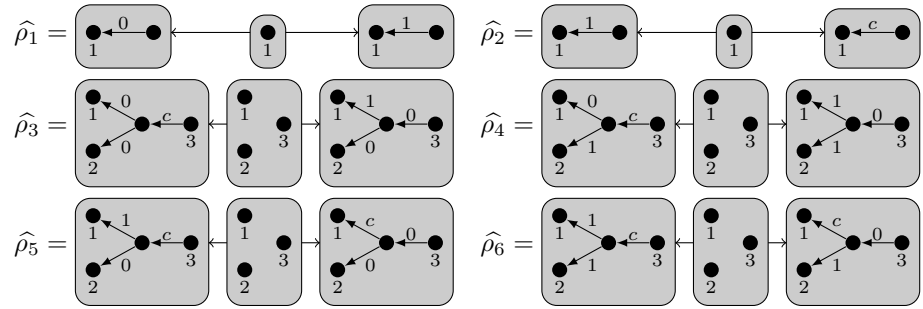
\begin{align*}
  \widehat{\rho_{1}} &= \ExpRuleInitTreeOne
  &\widehat{\rho_{2}} &= \ExpRuleInitTreeTwo
  \displaybreak[0] \\
  \widehat{\rho_{3}} &= \ExpRuleOneTree
  &\widehat{\rho_{4}} &= \ExpRuleTwoTree
  \displaybreak[0] \\
  \widehat{\rho_{5}} &= \ExpRuleThreeTree
  &\widehat{\rho_{6}} &= \ExpRuleFourTree
\end{align*}
  
The rules $\widehat{\rho_{1}}$ and $\widehat{\rho_{2}}$ increment the
shared least significant bit by $1$. These two rules can only be
applied at the root of the tree (due to the dangling edge condition of
the DPO approach), as long as the edge is either labelled~$0$ or~$1$.
By applying the rules $\widehat{\rho_{3}},\dots,\widehat{\rho_{6}}$, a
carrier bit can be passed to the next bit.  Proving termination of
this graph transformation system is non-trivial. By applying for
instance $\widehat{\rho_{6}}$, the value of the counters containing
interface node~$1$ does not change, while other counter values decrease.

\begin{wrapfigure}{r}{0.44\textwidth}
  \vspace{-1.0cm}
  \begin{tabular}{rl}
    $\widehat{T} = {}$ &
    \begin{tikzpicture}[x=1.5cm,y=-1.2cm,baseline=(1.south)]
      \node[gnode] (1) at (0,0) {} ;
      \node[gnode] (2) at (1.5,0) {} ;
      \draw[gedge] (1) .. controls +(70:1cm) and +(110:1cm) .. 
                   node[arlab,above] {$1^1$} (1) ;
      \draw[gedge] (1) .. controls +(160:1cm) and +(200:1cm) .. 
                   node[arlab,left] {$0^1$} (1) ;
      \draw[gedge] (1) .. controls +(250:1cm) and +(290:1cm) .. 
                   node[arlab,below] {$c^1$} (1) ;
      \draw[gedge] (2) to[bend left=20] node[arlab,below] {$1^1$} (1) ;
      \draw[gedge] (2) to[bend right=20] node[arlab,above] {$0^2$} (1) ;
      \draw[gedge] (2) .. controls +(70:1cm) and +(110:1cm) .. 
                   node[arlab,above] {$1^2$} (2) ;
      \draw[gedge] (2) .. controls +(340:1cm) and +(20:1cm) .. 
                   node[arlab,right] {$0^2$} (2) ;
      \draw[gedge] (2) .. controls +(250:1cm) and +(290:1cm) .. 
                   node[arlab,below] {$c^2$} (2) ;
    \end{tikzpicture}
  \end{tabular}
  \vspace{-.9cm}
\end{wrapfigure}

\noindent We evaluate the rules with respect to the following weighted
type graph $\widehat{T}$ over the arithmetic semiring.  We can prove
that $\widehat{\rho_{1}}$ and $\widehat{\rho_{2}}$ are decreasing and
$\widehat{\rho_{3}},\dots,\widehat{\rho_{6}}$ are non-increasing,
which means that $\widehat{\rho_{1}}$, $\widehat{\rho_{2}}$ can be
removed using a relative termination argument.

The rules $\widehat{\rho_{3}}$ and $\widehat{\rho_{4}}$ can be removed due to 
the decreasing number of c-labelled edges, which remain constant
in $\widehat{\rho_{5}}$ and $\widehat{\rho_{6}}$. Afterwards we can remove 
$\widehat{\rho_{5}}$, $\widehat{\rho_{6}}$ since they decrease the number of 
1-labelled edges. The graph transformation system has been shown to terminate 
uniformly, since there are no rules left.

\section{Finding Weighted Type Graphs and Implementation}
\label{sec:implementation}

The question of how to find suitable weighted type graphs has been
left open so far. Instead of manually searching for a suitable type
graph we employ a satisfiable modulo theories (SMT) solver (in this
case Z3) that can solve inequations over the natural numbers.

We fix a number $n$ of nodes in the type graph and proceed as follows:
take a complete graph $T$ with $n$ nodes, i.e., a graph with an edge
for every pair $i,j\in \{1,\dots,n\}$ of nodes and every edge label
$a\in\Lambda$. Every edge $e$ in this graph is associated with a
variable $x_e$. The task is to assign weights to those variables such
that rules can be shown as either decreasing or non-increasing.

Now, for every rule $\rho = L\arleft[\phi_L] I \arright[\phi_R] R$ and
every map $t\colon I\to T$ we obtain an inequation:
\[ \sum_{\substack{t_L\colon L \to T\\ t_L \circ \phi_L = t}}
\prod_{e\in E_L} x_{t_L(e)} \ge \sum_{\substack{t_R\colon R \to T\\
    t_R \circ \phi_R = t}} \prod_{e\in E_R} x_{t_R(e)} \] If we want
to show that $\rho$ is decreasing and $t$ is the flower morphism $\ge$
has to be replaced by $>$.

Doing this for each rule and every map $t$ gives us equations that can
be used as input for an SMT-solver. We consider the weights as natural
numbers only up to a given bound by restricting the length of the
corresponding bit-vectors. Note that we would be outside the decidable
fragment of arithmetics otherwise since the equations would contain
multiplication of variables (as opposed to multiplication of constants
and variables). By using a bit-vector encoding the SMT-solver Z3 can
reliably find a solution (if it exists) and especially such solutions
are found for the examples discussed in Section~\ref{sec:example}. Any
solution gives us a valid weighted type graph.

A prototype Java-based tool, called \emph{Grez}, has been written and
was introduced in \cite{bkz14}. Given a graph transformation system
$\mathcal{R}$, the tool tries to automatically find a proof for the
uniform termination of $\mathcal{R}$. The tool supports relative
termination and runs different algorithms (which are chosen by the user) 
concurrently to search a proof. If one algorithm succeeds in finding a 
termination argument for at least one of the rules, all processes are 
interrupted and the corresponding rule(s) will be removed from $\mathcal{R}$. 
The algorithms are then executed on the smaller set of rules and this
procedure is repeated until all rules have been removed. Afterwards
\emph{Grez} generates the full proof which can be saved as a PDF-file.

\emph{Grez} provides both a command-line interface and a graphical
user interface. The tool supports the integration of external tools,
such as other termination tools or SMT-solvers. \emph{Grez} can use
any SMT-solver which supports the SMT-LIB2 format \cite{bst10}.
\emph{Grez} generates the inequation described above in this format
and passes it, either through a temporary file or via direct output
stream, to the SMT-solver. The results are parsed back into the
termination proof, as soon as the SMT-solver terminates and produces a
model for the formula.

We ran the tool on all examples of this paper using a Windows
workstation with a $2,67$ Ghz, 4-core CPU and $8$ GB RAM. All proofs
were generated in less than $1$ second. The tool, a user manual
\cite{bru15} and the examples from this paper can be downloaded from
the \emph{Grez} webpage\short{:
  www.ti.inf.uni-due.de/research/tools/grez.}\full{.\footnote{\url{www.ti.inf.uni-due.de/research/tools/grez}}}

\section{Conclusion} 
\label{sec:conclusion}

We have shown how to extend the tropical and arctic weighted type
graphs of \cite{bkz14} to weighted type graphs over general semirings
and their application to the termination analysis of graph
transformation systems. This enables us to work in the arithmetic
semiring and to prove termination of systems that could not be handled
with previous approaches. Note that arithmetic type graphs do not
subsume previous termination analysis methods, but rather complement
them. In practice one should always try several methods in parallel
threads, as it is done in our termination tool Grez.

\smallskip

\noindent\emph{Related Work.} As already mentioned in the
introduction, there is some work on termination analysis for graph
transformation systems, often using rather straightforward counting
arguments. Some work is specifically geared to the analysis of model
transformations, taking for instance layers into account.

The paper \cite{bhpt05} considers high-level replacement units
(\textsc{hlru}), which are transformation systems with external
control expressions. The paper introduces a general framework for
proving termination of such \textsc{hlru}s, but the only concrete
termination criteria considered are node and edge counting, which are
subsumed by the weighted type graph method (for more details see
\cite{bkz14}).

In \cite{eeltvv05} layered graph transformation systems are
considered, which are graph transformation systems where interleaving
creation and deletion of edges with the same label is prohibited and
creation of nodes is bounded.  The paper shows such graph
transformation systems are terminating.

Another interesting approach encodes graph transformation systems into
Petri nets \cite{vvept06} by introducing one place for every edge
label and transforming rules into transitions. Whenever the Petri net
terminates on all markings, we can conclude uniform termination of the
original graph transformation rules. Note that the second example of
Section~\ref{sec:example} can not be handled in this way by Petri
nets.\footnote{Starting with three edges labelled
  $0,1,\mathit{count}$, rule $\rho'_2$ transforms them into three
  labels $0,c,\mathit{count}$, which, via rule $\rho'_3$, are again
  transformed into $0,1,\mathit{count}$.} On the other hand
\cite{vvept06} can handle negative application conditions in a limited
way, a feature we did not consider here.

Another termination technique via forward closures is presented in
\cite{p:termination-graph-rewriting}. Note that the example discussed
in this paper (termination of a graph transformation system based on
the string rewriting rules $ab\to ac, cd\to db$) can be handled by our
tool via tropical type graphs.

\smallskip

\noindent\emph{Future Work.} Naturally, integration of (negative) 
application condition is an interesting direction for future work.
Furthermore we have already started to work on techniques for pattern
counting. Here we are interested in deciding, whether a given rule
$\rho$ always decreases the number of occurrences of a given subgraph
$P$.

Another area of future research that might be of great interest is
non-uniform termination analysis, i.e., to analyse whether the rules
terminate only on a restricted set of graphs. In applications it is
often the case that rules do not always terminate, but they terminate
on all input graphs of interest (lists, cycles, trees, etc.). For
this, it will be necessary to find a suitable way to characterize
graph languages that is useful for the application areas and
integrates well with termination analysis.

\bibliography{references}
\bibliographystyle{plain}

\full{
\appendix

\section{Proofs}
\label{sec:proofs}

\begin{lemma_for}{lem:med}{.}
  \lemmediating
\end{lemma_for}

\begin{proof}
  It is straightforward to verify that $\mathit{med}_{\mathit{PO}}$
  and $\mathit{med}^{-1}_{\mathit{PO}}$ are indeed inverse to each
  other and hence both are bijections. \qed
\end{proof}

\begin{lemma_for}{lem:properties-type-graph}{\ (Properties of weighted
    type graphs).}
  \propertiestypegraph
\end{lemma_for}
\begin{proof}~
  \begin{enumerate}[(i)]
  \item $\fFlowerM[T]{G}$ exists by construction.  Furthermore, since
    $w_T(e)\in S_<$ for all edges in the range of $\fFlowerM[T]{G}$,
    it holds that $\fFlowerM[T]{G}\in S_<$, using the fact that
    $1\in S_<$ (needed when $E_G=\emptyset$) and $S_<$ is closed under
    multiplication.
  \item
    Since $G_0$ is discrete and the square is a pushout, the edge
    set $E_G$ is (isomorphic to) the disjoint union of $E_{G_1}$ and
    $E_{G_2}$. Thus:
    \begin{align*}
      w_T(t) &= \prod_{\mathclap{e\in E_G}} w_T(t(e)) =
      \prod_{\mathclap{e\in E_{G_1}}} w_T((t \circ \phi_1)(e)) \otimes
      \prod_{\mathclap{e\in E_{G_2}}} w_T((t \circ \phi_2)(e)) \\
      &= w_T(t \circ \phi_1) \otimes w_T(t \circ \phi_2),
    \end{align*}
    as required.
    \qed
  \end{enumerate}
\end{proof}

\begin{lemma_for}{lem:decreasing}{.}
  \lemdecreasing
\end{lemma_for}

\begin{proof}
  Let $\rho = L\arleft[\phi_L] I \arright[\phi_R] R$.  The rewriting
  step $G \Rightarrow_\rho H$ is depicted below on the left.

  For every possibility to type $G$ via $t_G\colon G\to T$, there
  exists a morphism $t_C = t_G\circ \psi_L\colon C\to T$ and we obtain
  $t_H\colon H\to T$ as mediating morphism of the right-hand pushout
  $\mathit{PO}_2$ (see diagram on the right).
  \begin{center}
    \begin{tikzpicture}[x=1.5cm,y=-1.5cm,baseline=(i)]
      \node (l) at (-1,0) { $L$ } ;
      \node (i) at ( 0,0) { $I$ } ;
      \node (r) at ( 1,0) { $R$ } ;
      \node (g) at (-1,1) { $G$ } ;
      \node (c) at ( 0,1) { $C$ } ;
      \node (h) at ( 1,1) { $H$ } ;
      \draw[->] (i) -- node[arlab,above] {$\phi_L$} (l) ;
      \draw[->] (i) -- node[arlab,above] {$\phi_R$} (r) ;
      \draw[->] (l) -- node[arlab,left] {$m$} (g) ;
      \draw[->] (i) -- node[arlab,left] {$c$} (c) ;
      \draw[->] (r) -- node[arlab,left] {$n$} (h) ;
      \draw[->] (c) -- node[arlab,above] {$\psi_L$} (g) ;
      \draw[->] (c) -- node[arlab,above] {$\psi_R$} (h) ;
    \end{tikzpicture}
    \qquad
    \begin{tikzpicture}[x=1.5cm,y=-1.5cm,baseline=(i)]
      \node (po1) at (-.5,.5) { $\mathit{PO}_1$ } ;
      \node (po2) at ( .5,.5) { $\mathit{PO}_2$ } ;
      \node (l) at (-1,0) { $L$ } ;
      \node (i) at ( 0,0) { $I$ } ;
      \node (r) at ( 1,0) { $R$ } ;
      \node (g) at (-1,1) { $G$ } ;
      \node (c) at ( 0,1) { $C$ } ;
      \node (h) at ( 1,1) { $H$ } ;
      \node (t) at ( 0,2) { $T$ } ;
      \draw[->] (i) -- node[arlab,above] {$\phi_L$} (l) ;
      \draw[->] (i) -- node[arlab,above] {$\phi_R$} (r) ;
      \draw[->] (l) -- node[arlab,left] {$m$} (g) ;
      \draw[->] (i) -- node[arlab,left] {$c$} (c) ;
      \draw[->] (r) -- node[arlab,left] {$n$} (h) ;
      \draw[->] (c) -- node[arlab,above] {$\psi_L$} (g) ;
      \draw[->] (c) -- node[arlab,above] {$\psi_R$} (h) ;
      \draw[->] (g) -- node[arlab,below left] {$t_G$} (t) ;
      \draw[->] (c) -- node[arlab,left,near start] {$t_C$} (t) ;
      \draw[->] (l) to[out=-145,in=-180] node[arlab,left] {$t_L$} (t) ;
      \draw[->] (r) to[out=-45,in=0] node[arlab,right] {$t_R$} (t) ;
      \draw[->,dashed] 
                (h) -- node[arlab,below right] {$t_H$} (t) ;
    \end{tikzpicture}
  \end{center}
  Now we have (compare with the diagram above on the right):
  \begin{align}
    w_T(G) &= \sum_{t_G\colon G\to T} w(t_G)
    \\     &= \sum_{t_C\colon C\to T} 
              \sum_{\substack{t_L\colon L\to T\\
                              t_L\circ \phi_L = t_C\circ c}} 
                          w(\mathit{med}_{\mathit{PO}_1}(t_C,t_L))
              \label{eq:use-po}
    \\     &= \sum_{t_C\colon C\to T} 
              \sum_{\substack{t_L\colon L\to T\\
                              t_L\circ \phi_L = t_C\circ c}} 
                          \bigl( w(t_L) \otimes w(t_C) \bigr)
              \label{eq:use-stable}
    \\     &= \sum_{t_C\colon C\to T} 
              \Bigl( w(t_C) \otimes
                     \sum_{\substack{t_L\colon L\to T\\
                                     t_L\circ \phi_L = t_C\circ c}} 
                       w(t_L)
              \Bigl)
              \label{eq:use-distr}
    \\     &= \sum_{t_C\colon C\to T} 
              \bigl( w(t_C) \otimes w_{t_C\circ c}(\phi_L) \bigr)
              \label{eq:use-def}
  \end{align}              
  where (\ref{eq:use-po}) follows from the fact that $\mathit{med}$ is
  a bijection (see Lemma~\ref{lem:med}), (\ref{eq:use-stable}) is an
  application of the equation $w(t) = w(t\circ\phi_1) \otimes
  w(t\circ\phi_2)$ of Lemma~\ref{lem:properties-type-graph},
  (\ref{eq:use-distr}) follows from distributivity and
  (\ref{eq:use-def}) holds by definition.  Symmetrically, we have
  \[
    w_T(H) = \sum_{t_C\colon C\to T} 
             \bigl( w(t_C) \otimes w_{t_C\circ c}(\phi_R) \bigr).
  \]
  Using this, we can prove the two parts of the lemma.
  
  \begin{enumerate}[(i)]  
  \item Since $\rho$ is non-increasing, it holds by definition that
    $w_{t_C\circ c}(\phi_L) \geq w_{t_C\circ c}(\phi_R)$ for all
    $t_C\colon C\to T$. Using the fact that all weights in the type
    graph are contained in $S_\le$ and $w_T(t_C)$ is obtained by
    multiplying such weights, we can infer that:
    \[ w(t_C)\otimes w_{t_C\circ c}(\phi_L) \geq w(t_C) \otimes
      w_{t_C\circ c}(\phi_R). \] From that it follows that
    $w_T(G) \geq w_T(H)$.
  \item Since $\rho$ is decreasing, it additionally holds by
    assumption that
    $w_{\fFlowerM{I}}(\phi_L) > w_{\fFlowerM{I}}(\phi_R)$.  Since
    $w(\fFlowerM{C}) \in S_<$
    (cf. Definition~\ref{def:weighted-type-graph} and closure of $S_<$
    under multiplication), by Lemma~\ref{lem:properties-type-graph},
    we have that
      \[
        w(\fFlowerM{C}) \otimes w_{\fFlowerM{I}}(\phi_L) > 
        w(\fFlowerM{C}) \otimes w_{\fFlowerM{I}}(\phi_R).
      \]
      The latter gives us the summands on both sides for the case of
      $t_C=\fFlowerM{C}$ and $t_C\circ c = \fFlowerM{I}$.  Together
      with the inequalities from~(i) it follows that $w_T(G) > w_T(H)$
      (using the fact that $S$ is a strictly ordered semiring).  \qed
    \end{enumerate}
\end{proof}

\begin{theorem_for}{thm:terminating}{.}
  \thmterminating
\end{theorem_for}

\begin{proof}
  ($\Rightarrow$): 
    It is an immediate consequence of $R$ being terminating that its subset
    $R^{=}$ is also terminating.
   
  ($\Leftarrow$):
    For a rule $\rho$ and transition $G \Rightarrow_r H$, it holds,
    by Lemma~\ref{lem:decreasing}, that $w_T(G) > w_T(H)$ if 
    $\rho\in R^{<}$ and $w_T(G) \geq w_T(H)$ if $\rho\in R^{=}$.
    From this it follows that each infinite transition sequence
    of $R$ ends in an infinite transition sequence of $R^{=}$,
    which do not exist by assumption. \qed
\end{proof}

\begin{lemma_for}{lem:decreasing-nonstrict}{.}
  \lemdecreasingnonstrict
\end{lemma_for}

\begin{proof}
  The proof proceeds analogously to the proof of
  Lemma~\ref{lem:decreasing}. For non-increasing rules the proof is
  exactly the same.

  For strongly decreasing rules we have to show 
  \[ \sum_{t_C\colon C\to T} 
  \bigl( w(t_C) \otimes w_{t_C\circ c}(\phi_L) \bigr) >
  \sum_{t_C\colon C\to T} 
  \bigl( w(t_C) \otimes w_{t_C\circ c}(\phi_R) \bigr)
  \] 
  This holds since $w_{t_C\circ c}(\phi_L) > w_{t_C\circ c}(\phi_R)$
  for all $t_C$, and $w(t_c)\in S_<$ (due to the fact that all weights
  in the type graph are in $S_<$ and $S_<$ is closed under
  multiplication). Hence the properties of strongly ordered semirings
  allow us to conclude. \qed
\end{proof}
}

\end{document}